\documentclass[twocolumn]{aastex631}

\usepackage[T1, T2A]{fontenc}
\usepackage[utf8]{inputenc}

\shorttitle{Ultra-short-period contact eclipsing binaries}
\shortauthors{Papageorgiou et al.}
\graphicspath{{./}{figures/}}
\begin{document}

\title{Three Ultra-short Period Contact Eclipsing Binary Systems Mined from Massive Astronomical Surveys}

\correspondingauthor{Athanasios Papageorgiou}
\email{apapageorgiou@upatras.gr}
 
\author[0000-0002-3039-9257]{Athanasios Papageorgiou}
\affiliation{Department of Physics, University of Patras, 26500, Patra, Greece}

\author{Panagiota-Eleftheria Christopoulou}
\affiliation{Department of Physics, University of Patras, 26500, Patra, Greece}

\author{C. E. Ferreira Lopes}
\affiliation{Instituto de Astronomía y Ciencias Planetarias, Universidad de
Atacama, Copayapu 485, Copiap\'{o}, Chile}
\affiliation{Universidade de S\~{a}o Paulo, IAG, Rua do Mat\~{a}o 1226, Cidade Universit\'{a}ria, S\~{a}o Paulo, 05508-900, Brazil}
\affiliation{National Institute For Space Research (INPE/MCTI), Av. dos Astronautas, 1758 - S\~{a}o Jos\'{e} dos Campos - SP, 12227-010, Brazil}

\author{Eleni Lalounta}
\affiliation{Department of Physics, University of Patras, 26500, Patra, Greece}
%%\collaboration{6}{(AAS Journals Data Editors)}

\author[0000-0001-6003-8877]{M\'{a}rcio Catelan}
\affiliation{Instituto de Astrofísica, Pontificia Universidad Cat\'{o}lica de Chile,\\
Av. Vicu\~{n}a Mackenna 4860, 7820436 Macul, Santiago, Chile}
\affiliation{Millennium Institute of Astrophysics, Nuncio Monse\~{n}or Sotero Sanz 100,\\ Of. 104, Providencia, Santiago, Chile}
\affiliation{Centro de Astro-Ingeniería, Pontificia Universidad Católica de Chile, Av. Vicuña Mackenna 4860, 7820436 Macul, Santiago, Chile}

\author{Andrew J. Drake}
\affiliation{California Institute of Technology, 1200 East California Boulevard, Pasadena, CA 
91225, USA}

%%\author{ }
%%\altaffiliation{ }
%%\affiliation{TeXnology Inc.}

%%\author{Julie Steffen}
%%\affiliation{AAS Director of Publishing}
%%\affiliation{American Astronomical Society \\
%%1667 K Street NW, Suite 800 \\
%%Washington, DC 20006, USA}

%%\author{Magaret Donnelly}
%%\affiliation{IOP Publishing, Washington, DC 20005}

%% Note that the \and command from previous versions of AASTeX is now
%% depreciated in this version as it is no longer necessary. AASTeX 
%% automatically takes care of all commas and "and"s between authors names.

%% AASTeX 6.31 has the new \collaboration and \nocollaboration commands to
%% provide the collaboration status of a group of authors. These commands 
%% can be used either before or after the list of corresponding authors. The
%% argument for \collaboration is the collaboration identifier. Authors are
%% encouraged to surround collaboration identifiers with ()s. The 
%% \nocollaboration command takes no argument and exists to indicate that
%% the nearby authors are not part of surrounding collaborations.

%% Mark off the abstract in the ``abstract'' environment. 
\begin{abstract}
We present the photometric analysis of three ultra-short period total eclipsing binaries in contact configuration, CRTS$\_$J172718.0+431624, OGLE-BLG-ECL-000104, and OGLE-BLG-ECL-000012, mined from massive astronomical surveys. Using the available archival light curves (LCs) from Vista Variables in the Via L\'{a}ctea (VVV), Optical Gravitational Lensing Experiment (OGLE), Zwicky Transient Facility (ZTF) and Catalina Sky Survey (CSS) in different passbands and new multi-band photometric observations with the 2.3\,m Aristarchos telescope at Helmos Observatory, their relative physical parameters were derived. We explored the parameter space by using  PIKAIA genetic algorithm optimizer. The best photometric solution and error budget estimation was adopted for each system through MCMC sampling of the global optimum.
The approximate absolute parameters were derived for each contact system adopting an empirical mass-luminosity relation. All three systems have a mass ratio lower than 0.5. The exchange between primary and secondary depths of CRTS$\_$J172718.0+431624 during the years 2016-2022 may be due to spot activity.    
In addition, we present a detailed analysis of the first well characterized shortest period contact eclipsing binary with total eclipses known so far (OGLE-BLG-ECL-000104). Thanks to VVV and OGLE LCs, new distances were derived for OGLE-BLG-ECL-000104 and OGLE-BLG-ECL-000012 using empirical period-luminosity relations. The origin and evolutionary status of all three ultra-short period contact binaries are thoroughly discussed in the context of the detached-binary formation channel.

\end{abstract}

%% Keywords should appear after the \end{abstract} command. 
%% The AAS Journals now uses Unified Astronomy Thesaurus concepts:
%% https://astrothesaurus.org
%% You will be asked to selected these concepts during the submission process
%% but this old "keyword" functionality is maintained in case authors want
%% to include these concepts in their preprints.
\keywords{ methods: data analysis --- binaries: eclipsing ---
stars: fundamental parameters}

%% From the front matter, we move on to the body of the paper.
%% Sections are demarcated by \section and \subsection, respectively.
%% Observe the use of the LaTeX \label
%% command after the \subsection to give a symbolic KEY to the
%% subsection for cross-referencing in a \ref command.
%% You can use LaTeX's \ref and \label commands to keep track of
%% cross-references to sections, equations, tables, and figures.
%% That way, if you change the order of any elements, LaTeX will
%% automatically renumber them.
%%
%% We recommend that authors also use the natbib \citep
%% and \citet commands to identify citations.  The citations are
%% tied to the reference list via symbolic KEYs. The KEY corresponds
%% to the KEY in the \bibitem in the reference list below. 

\section{Introduction} \label{sec:intro}
The period distribution of contact binaries (EWs) has a sharp lower limit, $P= 0.22$~days, that was pointed out a long time ago \citep{1992Rucinski,2007Rucinski,2011Norton}. During the last ten years, due to the plethora of new EWs  emerged from recent surveys this limit is diminishing to $\sim$ 0.2~days or less, but the sharp decline at 0.22~days remains \citep{Li2019}. 

\cite{adea14}  confirmed the existence of 15 candidates with EW light curve type  with $P< 0.212$~days using data from the Catalina Sky Survey (CSS), whereas \cite{2015AcA....65...39S} identified two stars with EW light curve types having $P< 0.19$~days in the Optical Gravitational Lensing Experiment \citep[OGLE; ][]{1992AcA....42..253U} fields toward the Galactic bulge (OGLE-III-IV). Recently the Zwicky Transient Facility (ZTF) classified $\sim$ 21000 potential EWs with a shorter cutoff period of 0.19~days \citep{2020ChenZTF}, and \cite{2021Bienias} presented the shortest-period system found in the $\it{Kepler}$ field with a period of 0.1855~d, in addition to six more with periods below 0.22~days.  Similarly, the first Transiting Exoplanet Survey Satellite (\textsc{TESS}) catalogue of eclipsing binaries \citep{2022ApJS..258...16P} lists 4584 objects, of which only 8 have periods less than 0.23~days and morphology parameters corresponding to contact binaries. Three of them are known EWs. The shortest-period TESS candidate has $P=0.1485855$~days. Moreover, orbital periods as low as $0.112-0.116$~days have been reported in near-contact  M-dwarf binary systems   \citep{2012Nefs,2022Koen}. 

All the above identifications of systems reported as candidate ultra-short period contact binaries (USPCBs) are subject to the variable star classification criteria of each survey. In many cases the selection criteria are based only on reddening and color indices, on photometry through only one or two filters, and on preliminary modeling. In the absence of radial velocity measurements, this may be more complicated if the binary inclination angle is low and the light curve morphology is nearly sinusoidal (partial eclipses). Until now, only a couple of systems with P<0.22~days have radial velocities based on spectroscopic observations, namely GSC 01387-00475, with a period of 0.2178~days \citep{2008RucPrib}, and one red EW M dwarf, SDSS J001641-000925, with a period of 0.19856~days \citep{Davenport2013}. Another  candidate ultra-short period M dwarf binary, CSS J001242.4+130809 ($P=0.164086$~days), discovered by \cite{adea14}, although modeled as contact,  is considered by the authors uncertain since its classification is based only on a single set of spectral features. Nevertheless, and as noted by \cite{2014ApJ...790..157D}, when the M dwarfs are in the spectral range from M0V to M2.5V (as it is the case of SDSS J001641-000925), they  give rise to the same radial velocity amplitudes as a binary comprised of a main sequence star with a compact (white dwarf/subdwarf) companion. There are also around a dozen EWs with  $P< 0.22$~days with total eclipses mined from recent surveys with dedicated multiband observations (see Section~\ref{sec:Discussion}).

Why periods shorter than 0.22~days cannot be achieved easily, and why are USPCBs so rare? \cite{1992Rucinski} originally proposed that below this period both components become fully convective and dynamically unstable. In the detached to contact evolution of binaries scenario, tidal interaction is important in changing the orbit of a close detached binary. The degree of interaction  is critically dependent on the ratio of stellar radius  to the separation of stars \citep{1981A&A....99..126H}. For a given total angular momentum a detached close binary will reach an equilibrium stable state or the two stars will spiral in towards each other and coalesce. The equilibrium stable state is characterized by circular, coplanar and synchronized orbits of the stars  if more than three quarters  of the total angular momentum are in the form of orbital angular momentum \citep{1980A&A....92..167H}. Starting from such a progenitor \cite{2006Stepien} proposed a model where the dominating mechanisms of the close binary orbit evolution are the magnetic braking due to magnetized winds of the components and the mass transfer between them.
Briefly, the model assumes that mass transfer with mass ratio inversion occurs following the Roche lobe overflow (RLOF) of the massive component and shortly thereafter the contact configuration is reached. According to \cite{2006Stepien} and \cite{2012Stepien}, it takes too long for a detached low-mass binary to reach contact (through RLOF) by  angular momentum loss (AML) via magnetized winds, and periods shorter than $P = 0.22$~days cannot be achieved easily. This happens because the  estimated AML time scales required for isolated detached binaries (no tertiary components) with an initial primary mass of less than $~0.7 M_{\sun}$ and initial period 2.0 d are too long and therefore the RLOF will  not occur within the Hubble time. As a result, only short-period EWs  with total mass $>1-1.2 \,M_{\sun}$ (and slightly less for systems in globular clusters) should exist, as lower-mass stars which could evolve to shorter periods have not yet had enough time to do so. Alternatively, \cite{2012Jiang} suggested that low-mass contact binaries have a very short lifetime, as they become unstable during the mass transfer process to reach contact and merge. Nevertheless, \cite{2020ZhangQian} semi-empirically estimated a  period limit around 0.15~days, by studying the correlation among orbital periods, mass ratios, masses, and radii of 365 studied  contact binaries.
Another explanation to be considered is the presence of a third companion causing AML from the central eclipsing system during its evolution from birth to contact, through the Kozai mechanism \citep{2001Eggleton,2007Fabrycky}. This agrees with the idea that all contacts are in triple or higher-order systems \citep{2006PribulaRuci,2006AJ....132..650D}.

Hence, USPCBs around or below 0.22~days are of great interest, as their short period is connected to the origin and evolutionary channels of contact binaries.One limitation of finding and studying new USPCBs in the optical, is their position in the inner galactic disc and bulge. Although for more than two decades, the contribution of OGLE observations in the $I$ (and $V$) bands \citep{2015AcA....65...39S,2016AcA....66..405S} revealed USPCBs in the above regions, the deep near-IR $(0.9-2.5 \, \mu {\rm m})$ Vista Variables in the Vía L\'{a}ctea survey \cite[VVV,][]{2010NewA...15..433M,2011rrls.conf..145C,2013BAAA...56..153C} can reach through the dust and shed light on more distant USPCBs towards the inner regions of the Milky Way. VVV and its extension, VVVX \citep[VVV eXtended;][]{2018Minniti}, provide multi-epoch photometry in the $K_{s}$ band over a baseline of about ten years for a bulge region of 300 degrees$^{2}$ and part of the southern disk. The combined information that can be mined from OGLE ($V,I$ bands) and VVV ($K_{s}$ band) time-series data in overlapping sky regions is invaluable for variable star research in general, and eclipsing binary studies in particular.

In this paper, we present the first multi-band photometric analysis of   CRTS$\_$J172718.0+431624 (hereafter CRTSJ172718), reported as a USPCB system by \cite{2013A&A...549A..86L}, and  OGLE-BLG-ECL-000104 (hereafter OGLE104), similarly reported by \cite{2015AcA....65...39S}. An additional USPCB, OGLE-BLG-ECL-000012 (hereafter OGLE012), classified as a contact binary by one of us (A.P.) while preparing a catalogue of VVV eclipsing binaries, is also reported here for the first time (Papageorgiou et al., in prep.). In Section~\ref{sec:Observations}, we describe all the available observational data, both new and archival (from surveys), and in Section~\ref{sec:Analysis} we explain the procedures adopted to model the light curves (LCs). A method of estimating the absolute parameters, together with our results, are presented in Section~\ref{sec:Results}. Finally, in Section~\ref{sec:Discussion} we compare our derived parameters with those derived for other USPCBs in the literature and investigate their possible origin.

\section{Observations} \label{sec:Observations}
Our targets, are mined from CSS \citep{adea14}, OGLE-III and OGLE-IV \citep[][covering the years 2001–2013]{2016AcA....66..405S}, VVV and  the Zwicky Transient Facility
(ZTF) \citep{2019PASP..131a8003M}. 

\subsection{New photometric observations} \label{subsec:NewPhot}
New multi-color photometric observations of CRTSJ172718 were carried out on August 10 and 11, 2016, using the Ritchey-Chr\'{e}tien 2.3\,m Aristarchos telescope at Helmos Observatory, Greece. This telescope is equipped with a liquid nitrogen cooled Princeton Instruments VersArray 1024B CCD camera with 1024 $\times$ 1024 pixels. The pixel scale is 0.28$\arcsec$, resulting in an effective field of view of $4.8 \arcmin \times 4.8 \arcmin$. We used a set of Johnson-Cousins $BVRI$ filters with 150, 45, 30, and 60 seconds exposure times, respectively and obtained 77 images in B, 78 in V, 80 in R and 80 in I. 
Our fully automated pipeline \citep{2015ASPC..496..181P} that incorporates PyRAF \citep{2012ascl.soft07011S} and the Astrometry.net packages \citep{2010AJ....139.1782L}, was used for standard data reduction and  photometry. Image reduction included bias and flat-field correction while instrumental magnitudes were determined with aperture photometry. Information about the target and the reference stars is given in Table~\ref{tab:Tab1_stars}. The typical errors in the final differential magnitudes throughout the observing run are 4-5 mmag. From our $BVRI$ observations we determined one heliocentric time of light minimum and confirmed its short period.  

\subsection{Photometric observations from OGLE, VVV, ZTF and CSS} \label{subsec:ArchPhot}
The VVV \cite[][]{2010NewA...15..433M,2011rrls.conf..145C,2013BAAA...56..153C} ESO Public Survey, conducted on the 4.1\,m Visible and Infrared
Survey Telescope for Astronomy (VISTA), located at ESO Paranal Observatory, Chile, was carried out in five near-infrared bands ($Z$, $Y$, $J$, $H$, $K_s$) using the VISTA InfraRed CAMera (VIRCAM). VIRCAM is a $4\times 4$ array of Raytheon VIRGO IR detectors ($2048 \times 2048$ pixels) with a pixel scale of $0.34\arcsec$/pix. Data reduction, photometry and its calibration were done by the VISTA Data Flow System \citep{2004SPIE.5493..401E}. Recently, \cite{2020MNRAS.496.1730F} published the first variability catalog covering the entire VVV survey, comprising $\sim 45 \times 10^{6}$ variable star candidates, while \cite{Herpich2021} presented an analysis of previously known variable sources in the VVV footprint.

OGLE has monitored the Galactic bulge in $V$ and $I$ for 30~years, using the 1.3\,m Warsaw telescope, located at Las Campanas Observatory, Chile. LCs of OGLE104 and OGLE012 from OGLE-III and OGLE-IV \citep[covering the years 2001–2013;][]{2016AcA....66..405S} were retrieved from the OGLE Collection of Variable Stars.$\footnote{\url{https://ogledb.astrouw.edu.pl/~ogle/OCVS/index.php}}$ CSS includes the Catalina Schmidt Survey and the Mount Lemmon Survey in Tucson, Arizona, and the Siding Spring Survey in Siding Spring, Australia.
In this study, we use the photometry of CRTSJ172718 provided by the Catalina Surveys Data Release 2 (CSDR2).$\footnote{\url{http://nesssi.cacr.caltech.edu/DataRelease/}}$ The observations are taken unfiltered, and the magnitudes are transformed to an approximate $V$ magnitude \citep[$V_{\rm CSS}$,][]{2013ApJ...763...32D}.

Photometric data of CRTSJ172718 are also obtained from the ZTF survey database. ZTF is a 48-inch Schmidt telescope with a 47 $deg^{2}$ field of view \citep{2019PASP..131a8003M} scanning the entire northern sky. We use ZTF Data Release 12 (DR12)$\footnote{\url{https://irsa.ipac.caltech.edu/}}$ photometry in $g$ (1378 points), $r$ (1320 points) and $i$ (392 points)  filters.

\section{Light Curve modeling} \label{sec:Analysis}
\begin{figure*}[ht!]
\gridline{\fig{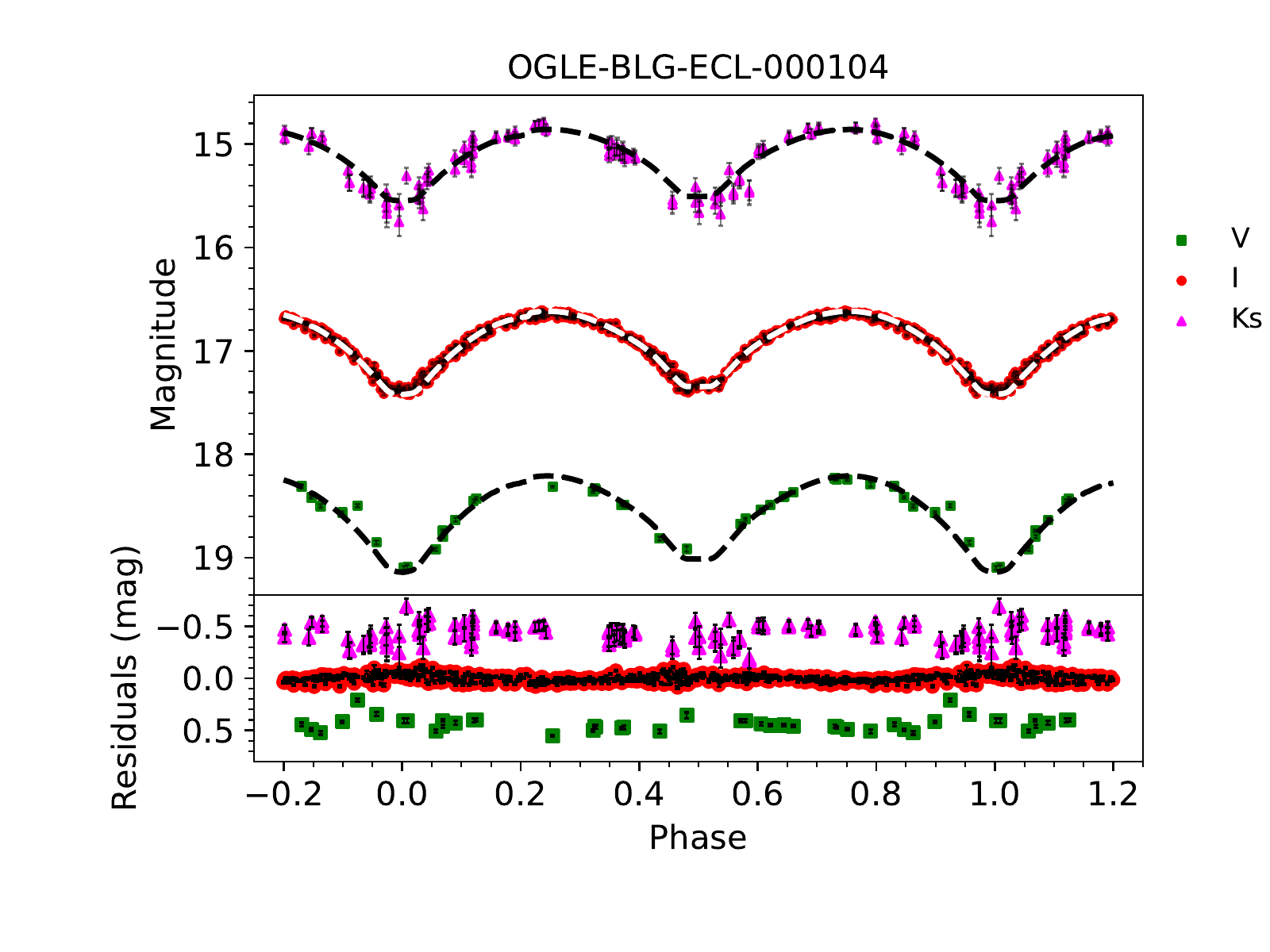}{0.5\textwidth}{(a) }
          \hspace{0.5cm}
          \fig{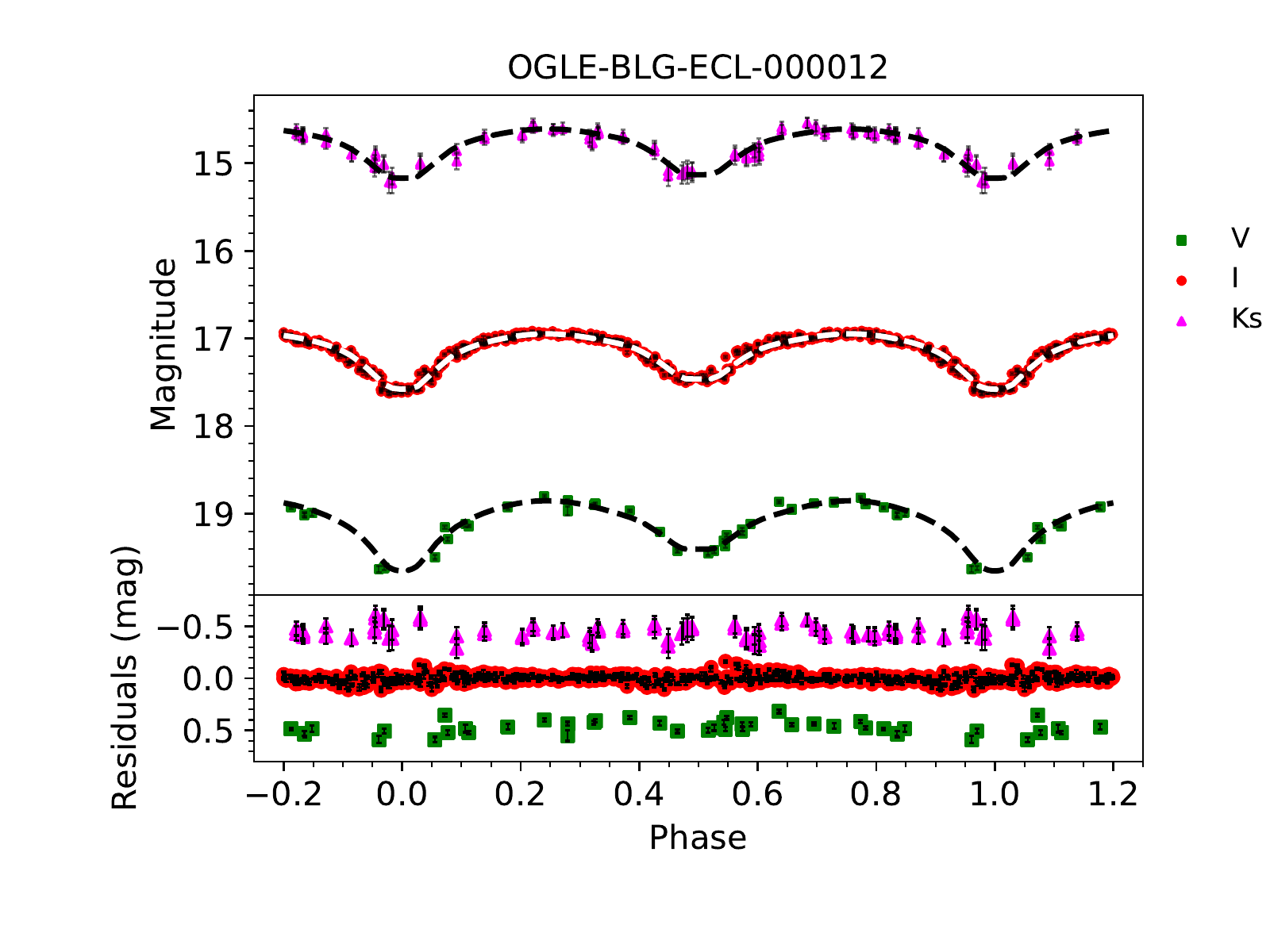}{0.5\textwidth}{(b) }}
\gridline{\fig{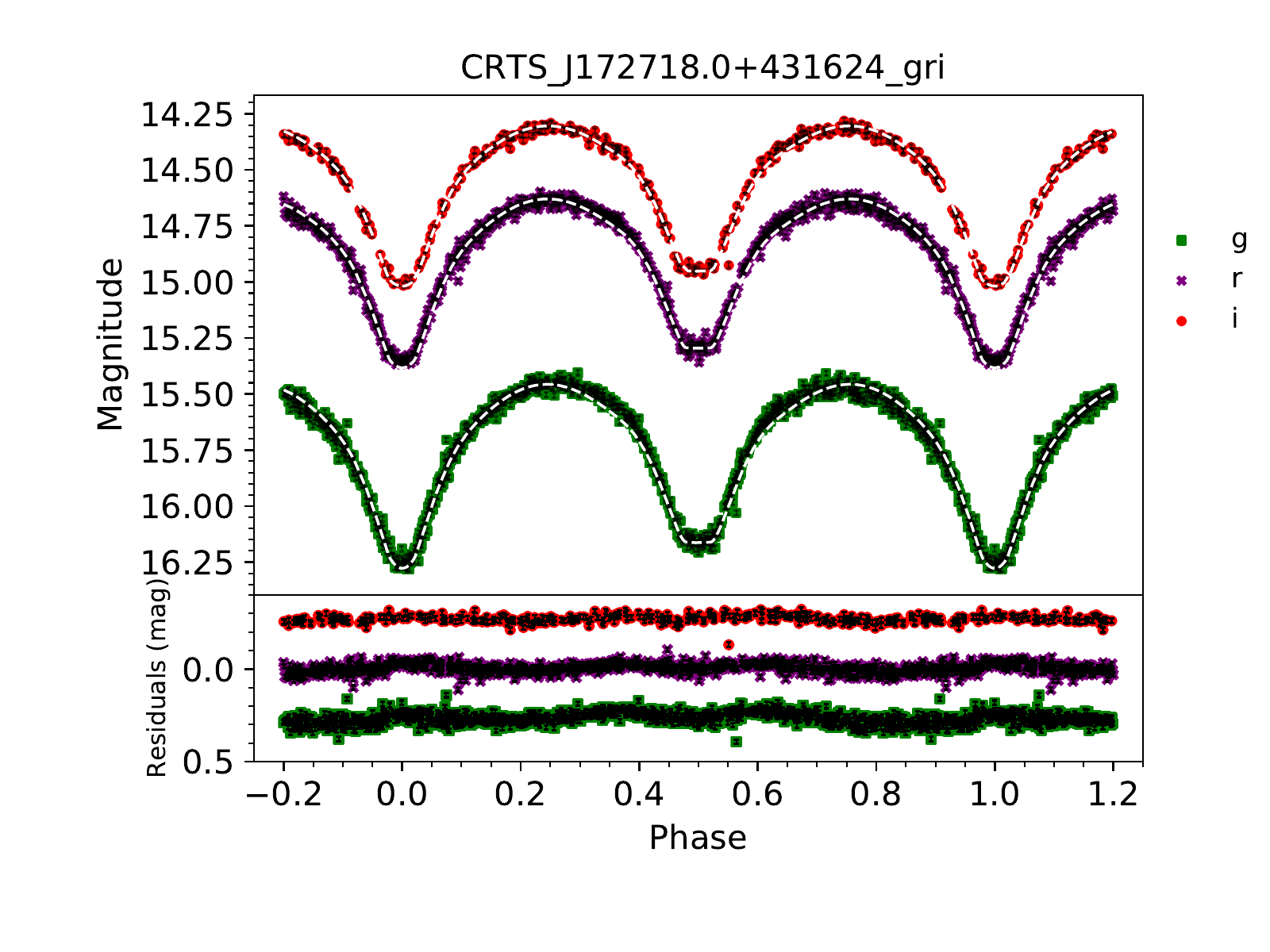}{0.5\textwidth}{(c)  }
          \fig{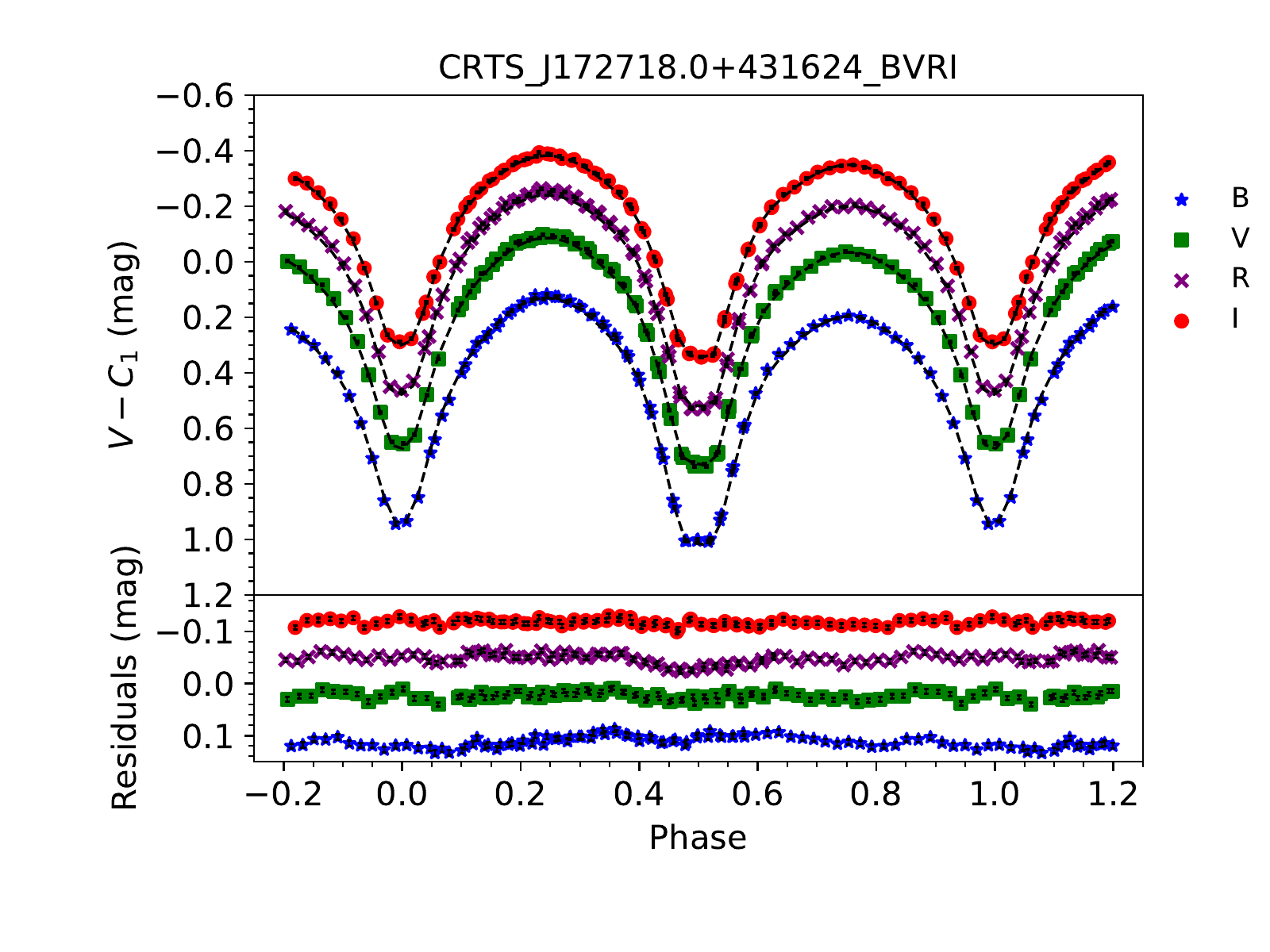}{0.5\textwidth}{(d) }}
\gridline{\fig{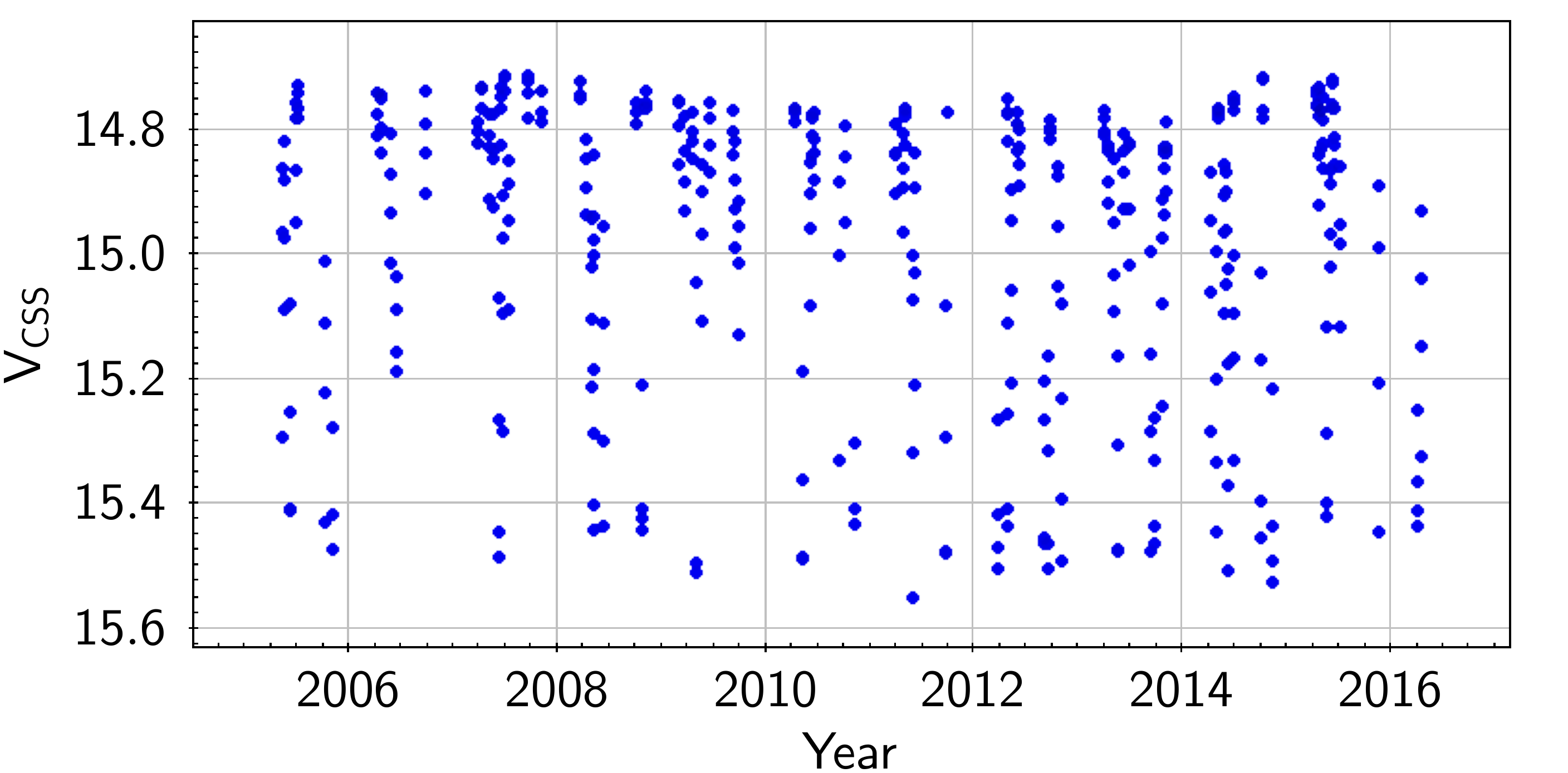}{0.5\textwidth}{(e)}}
\caption{Phase folded LCs for OGLE104 (panel a) and OGLE012 (panel b) in different bandpasses with error bars as shown by  green squares  ($V$), red dots ($I$) and purple triangles ($K_{s}$). Panel c-d: Phase folded  $gri$ and $BVRI$ LCs for CRTSJ172718 respectively as shown by blue stars ($B$), green squares ($g,V$), purple x marker ($r,R$), and red dots ($i,I$). In  panel (d) the LCs are shifted vertically for clarity, by the following amounts: $B-0.7$, $V-0.2$, $R+0.2$, $I+0.5$. The synthetic models (black or white dashed lines) from MCMC analysis are overplotted to the observed data. Bottom (panels a-d): Residuals between synthetic and photometric data. Panel e: The time series CSS data of CRTS$\_$J172718.0+431624 (2005-2016). }\label{fig:Lcs}
\end{figure*}

The selected USPCBs show total eclipses as shown in Figure~\ref{fig:Lcs}, which are more obvious in the I-band for OGLE104 and OGLE012. In the absence of radial velocity data, a reliable photometric mass ratio, $q=\frac{M_{2}}{M_{1}}$, can be derived using the photometric LCs with total eclipses \citep{2005Ap&SS.296..221T,2013CoSka..43...27H,2016PASA...33...43S}.  Based on the temperatures of the systems (see Section~\ref{sec:Results}), convective atmospheres were assumed, and the bolometric albedos and gravity darkening coefficients were set at values $A_{1}=A_{2}=0.5$ \citep{1973AcA....23...79R} and $g_{1}=g_{2}=0.32$ \citep{1967ZA.....65...89L}, respectively, for each system. The limb-darkening coefficients were interpolated from \cite{1993AJ....106.2096V} tables with a logarithmic law. 

As the effective temperature of the primary component (star eclipsed at phase zero) is an input parameter, for CRTSJ172718 it was set equal to the system's temperature as given by \cite{2019AJ....158..138S} and kept as a constant parameter during the modeling. For OGLE104, OGLE012, the color information from the LCs was conserved during the fitting \citep{10.1088/978-0-7503-1287-5}, and the temperature of the primary component was calculated from a temperature-color relation \citep{2010A&A...512A..54C}, assuming solar metallicity. Before applying such a relation, the observed color index was corrected for reddening using \textsc{mwdust} \citep{2016ApJ...818..130B,2019ApJ...887...93G}, and the distance\footnote{During the submission process of this paper, {\em Gaia}'s DR3 \citep{Creevey2022} was released, whereupon new distances were provided for both OGLE104 and OGLE012. The DR3 and EDR3 parallaxes are identical for both stars. Unfortunately, since these parallaxes are of low quality ($\frac{\varpi}{\sigma_{\varpi}}$<~10, where $\varpi$ is the parallax and $\sigma_{\varpi}$ its error, they are faint stars with $G~>~16$~mag, and are located in the Galactic plane, their distances are not considered reliable. As suggested by the {\em Gaia} team, to overcome these limitations, the use of near-infrared photometry would be very helpful, which is in accordance with the scope and the methodology of this paper.} obtained as described in Sections~\ref{subsec3:OGLE104},~\ref{subsec4:OGLE-BLG-ECL-00001}. This was performed for two colors, namely $V-K_s$ and $V-I$, and the mean value of the derived temperatures was finally adopted as the effective temperature of the primary component.

We perform a parameter search via genetic algorithm (GA) optimizer PIKAIA$\footnote{\url{http://n2t.net/ark:/85065/d70r9ntr}}$ interfaced with PHOEBE-0.31a scripter \citep{2005ApJ...628..426P} as adapted by \cite{PapageorgiouPhD}, followed by the Markov Chain Monte Carlo (MCMC) sampling of the global optimum. The ``Overcontact not in thermal contact'' mode of the PHOEBE-0.31a scripter \citep{2005ApJ...628..426P} was used, assuming circular orbits. For the case of ZTF photometry we use the 2015 version of the Wilson-Devinney binary star modeling (W-D) code$\footnote{\url{ftp://ftp.astro.ufl.edu/pub/wilson/lcdc2015/}}$ \citep{1971ApJ...166..605W, 2014ApJ...780..151W, 2008ApJ...672..575W, 1990ApJ...356..613W}. The observed LCs were weighted according to their mean errors and were solved simultaneously.
We use PIKAIA$\footnote{\url{http://n2t.net/ark:/85065/d70r9ntr}}$, a GA based numerical optimization technique inspired from the biological process of evolution by means of natural selection \citep{Charbonneau02anintroduction}. PIKAIA consists of the following steps a) create randomly set of models according to their limits (``population''), b) calculate cost function value from the model (PHOEBE-scripter/W-D) (``fitness''), c) select pairs of  parameter sets from the current population with the probability of a given set being selected made according to the fitness (``parents''), d) apply crossover and mutation and generate new parameter sets (``offspring''), e) Repeat steps c-d until the size of the generated population is equal to the previous population, then go to (b), f) evolve the initial sets and create generations. We created 120 sets of starting values from uniform distribution for the mass-ratio q, the modified surface potential of the components $\Omega_{1,2}$, the effective temperature ratio $T_{r}$, the inclination i and the passband luminosities were computed from PHOEBE scripter. The margins were set to [0.2-0.65] for q, [0.9–1.1] for  $T_{r}$, [$80\degr-90\degr$] for i, and [$\Omega_{in}-\Omega_{out}$] 
for $\Omega_{1,2}$. The population size was set to 120 individuals and at least 1000 generations were computed. The crossover probability was set to 0.85 and the adjustable one-point mutation rate was based on fitness. The best individuals from this list are used to get the best-fit parameters. The final parameters from this method are listed in Table~\ref {tab:Tab2_GA}.

We use the affine invariant Markov Chain Monte Carlo (MCMC) Ensemble sampler implemented in the EMCEE \citep{2013PASP..125..306F} Python package coupled with the W-D code.
The parameters that could be determined by using only photometric light curves of total eclipsing contact binaries are the parameters that have a more significant effect on the LC morphology \citep{2011AJ....141...83P} such as the amplitude of variation, eclipse depth difference, etc. Thus, the space of the parameter search was restricted to $q, i, T_{r}$, $\Omega_{1,2}$ and $L_{1}$.
The same margins as in GA search were used for the uniform priors of the parameters $q, i, T_{r}$, $\Omega_{1,2}$ and $L_{1}$ in the MCMC sampling.   

\begin{figure*}[ht!]
\gridline{\fig{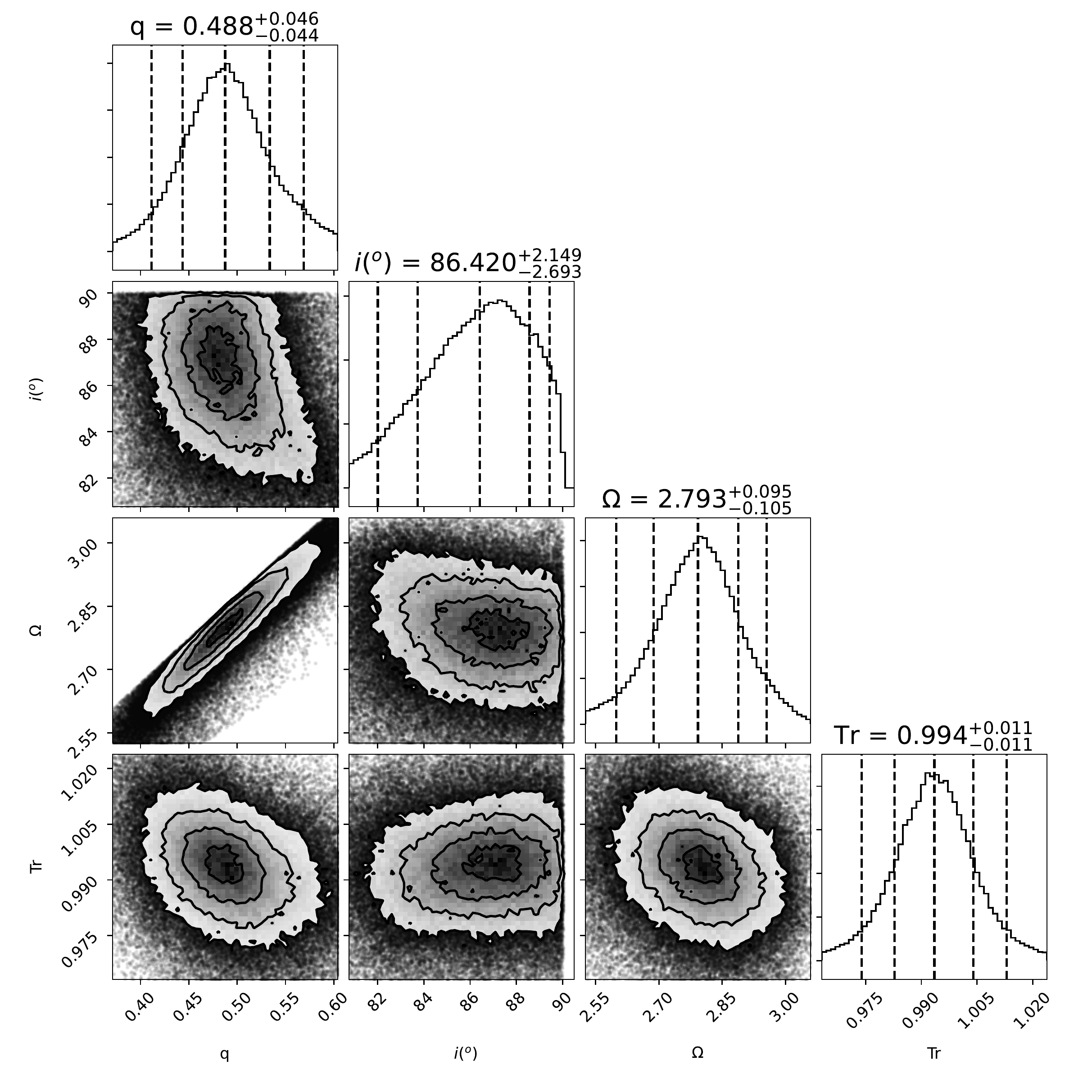}{0.5\textwidth}{(a)}
          \fig{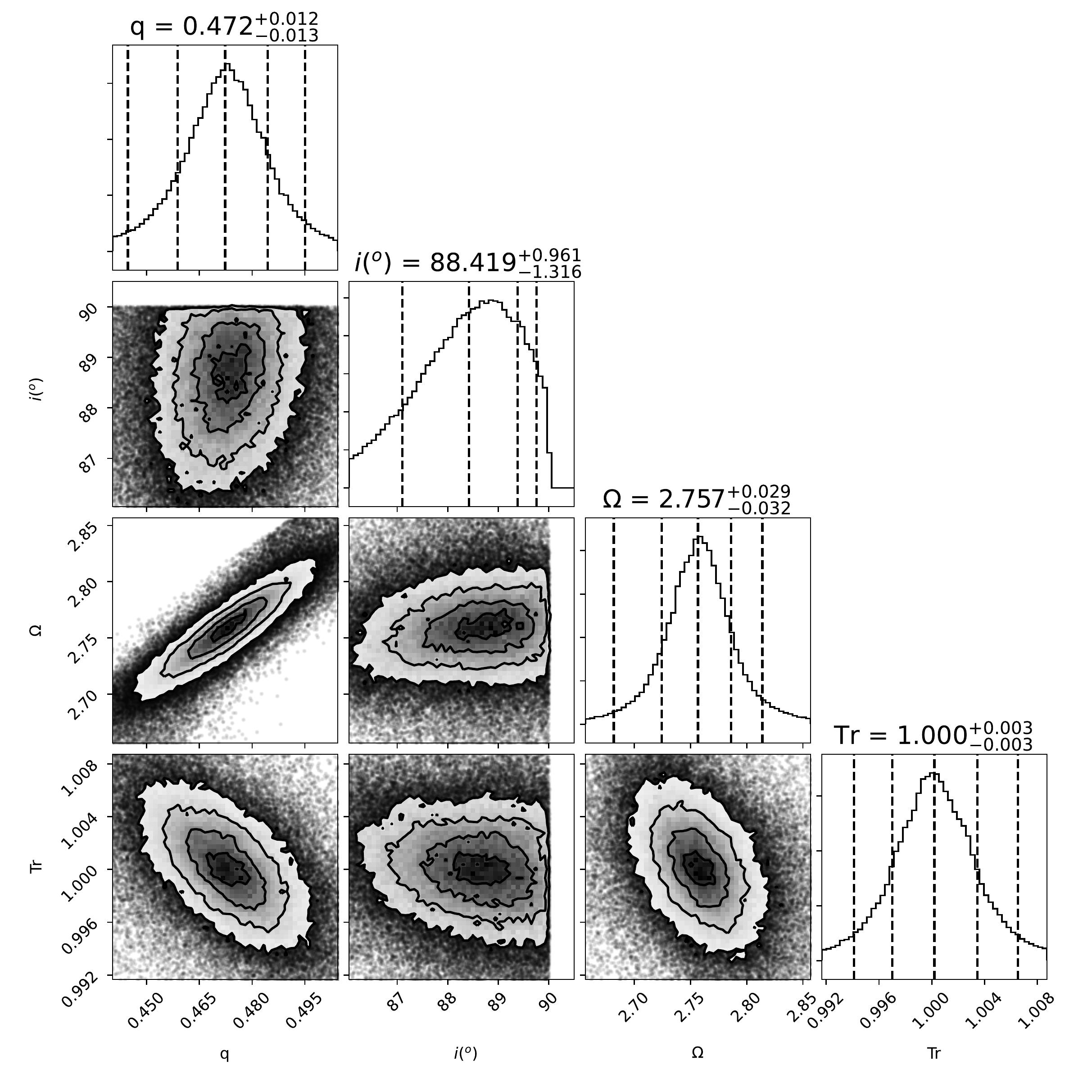}{0.5\textwidth}{(b)}}
\gridline{\fig{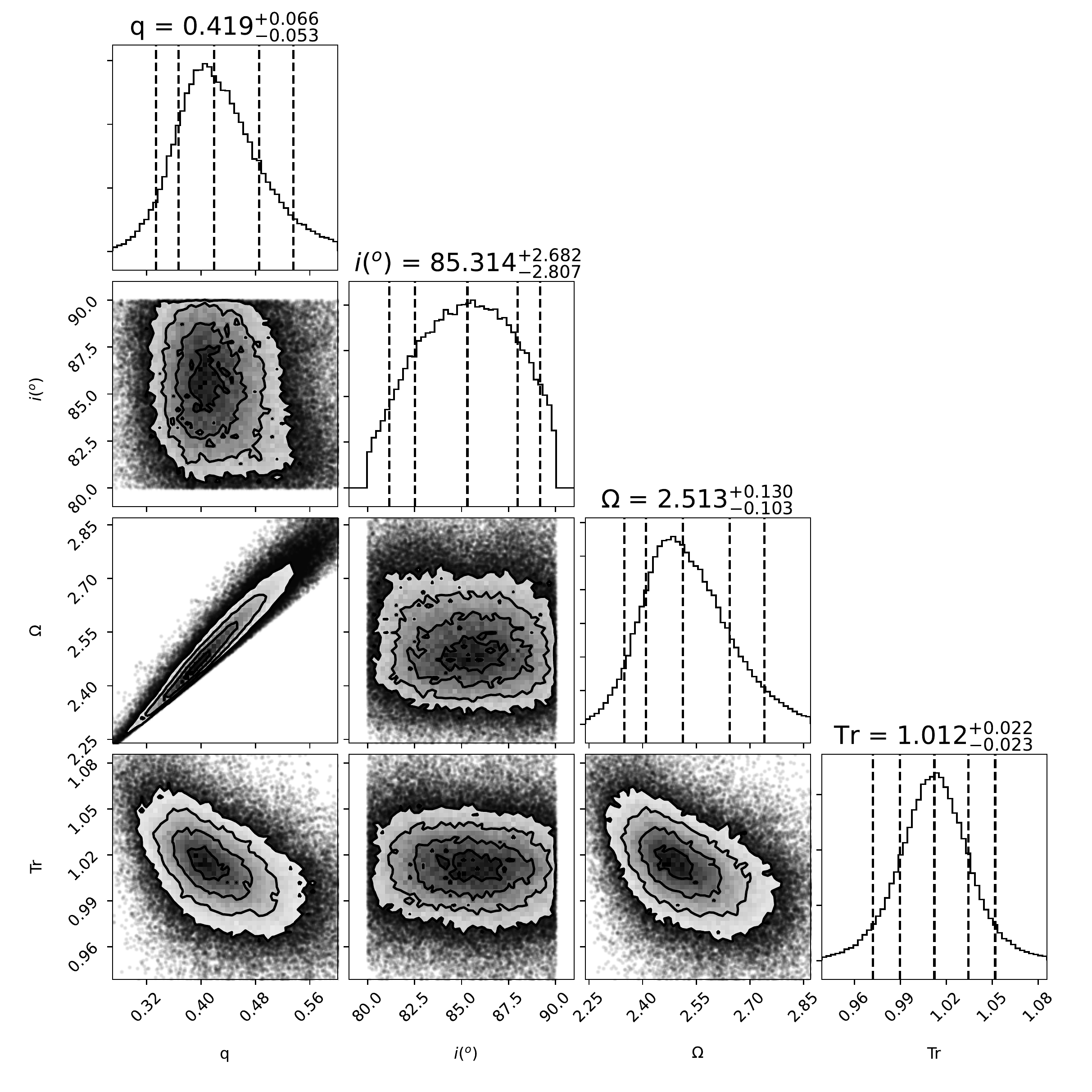}{0.5\textwidth}{(c)}
          \fig{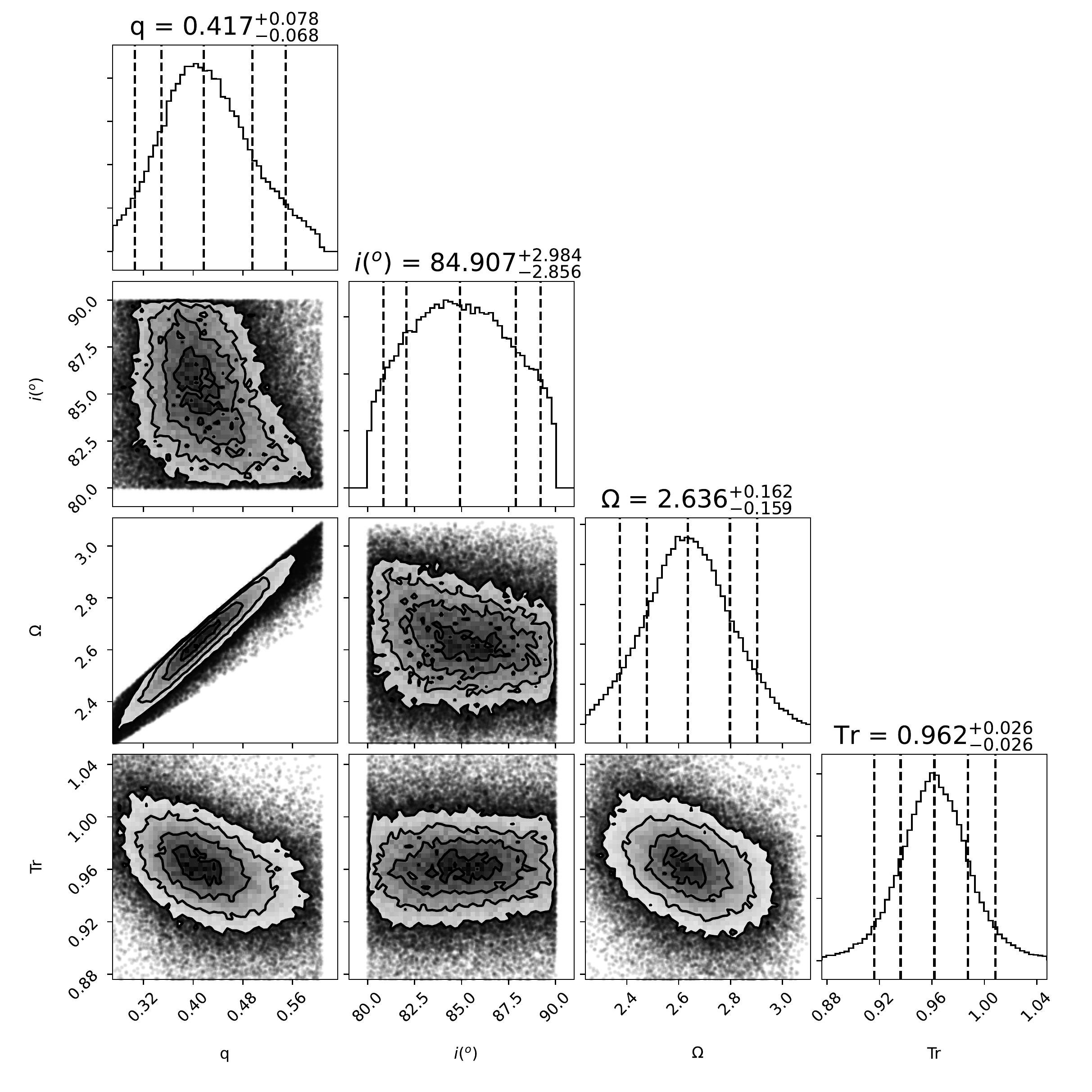}{0.5\textwidth}{(d)}
}
\caption{The probability distributions \citep{dan_foreman_mackey_2014_11020} of q, i, $\Omega$, and Tr determined by the MCMC modeling of CRTSJ172718 from $gri$ ZTF data (a) and our $BVRI$ data  (b), OGLE104 (c) and OGLE012 (d). }\label{fig:mcmc_plots}
\end{figure*}

A total of 30 walkers were used. The MCMC parameter search was run for 6,000-10,000 steps with a burn-in phase of 1,500-2,000 steps, for our $BVRI$ LCs of CRTSJ172718, and the three $V$, $I$, $K_s$  sets of LCs of OGLE104 and OGLE12, resulting in $\sim$180,000 iterations.
To achieve the MCMC chain convergence, several criteria were suggested for the chain(s) length, such as 10-20 times longer than the integrated autocorrelation time \citep{2020ApJS..250...34C, 2021ApJ...922..122L}. In order to secure convergence, we adopted the chain length to be longer than 50 times the integrated autocorrelation time\footnote{https://emcee.readthedocs.io/en/stable/tutorials/autocorr/}. The autocorrelation times for our MCMC chains were in the range of [1350-2560] iterations. Figures~\ref{fig:mcmc_plots}c-d show  the probability distributions of $q, i,  T_{r}$ and  $\Omega_{1,2}$ for OGLE104 and OGLE012 respectively. Posterior distributions indicate the probability that a given parameter set will describe the observed data. All of the posterior distributions are centralized and unimodal as shown in Figures~\ref{fig:mcmc_plots}c-d. The parameter uncertainties were derived from 1~$\sigma$ of the parameter posteriors.
The results are presented in Table~\ref{tab:Tab2} together with the coordinates ($RA_{J2000}$, $Dec_{J2000}$) and ephemerides. 

In the case CRTSJ172718  as displayed in Figure~\ref{fig:Lcs}d our 2016 LCs are  asymmetric showing differences in the maxima of 0.085-0.05 mag. However the newer $gri$ LCs of ZTF obtained in 2018-2022 show exchanges in depths of primary and secondary minima and no significant asymmetry at maxima. This made us to  search for an underlying model that can better constrain the solution and describe the physical parameters of the two components. The results of MCMC parameter search method with  W-D performed on the  ZTF data, using 10,000 steps with a 2,000 burn-in steps, resulting in 240,000 iterations are presented in Table~\ref{tab:Tab2} and in Figure~\ref{fig:mcmc_plots}a. We considered this model as the underlying model that can describe the system and explored the light curve solution of our 2016 data with a spotted model using a GA optimizer. Genetic algorithms show promise for determining spot parameters from simultaneously solution of LCs with broad wavelength coverage \citep{2022Galax..10....8T}. To investigate spot parameters, the PHOEBE program was interfaced with PIKAIA. In the first run we kept fixed q, i, $T_{r}$, $\Omega_{1,2}$ at their ZTF values and determine the spot parameters (longitude, latitude, spot size and temperature). In the second run, MCMC with W-D was used to adjust the parameters q, i, $T_{r}$, $\Omega_{1,2}$ and $L_{1}$ by keeping fixed the spot parameters (8,000 steps resulting in 240,000 iterations). The third light contribution was also investigated and found to be negligible ($\leq 1\%$) in the context of our photometric accuracy. Figure~\ref{fig:mcmc_plots}b shows the probability distributions of the above parameters of this spotted solution of our $BVRI$ data. All of the posterior distributions are centralized and unimodal as shown in Figures~\ref{fig:mcmc_plots}a-b. Only  the posterior distributions of the inclination are skewed left with the modes to higher values (88-89$\degr$). The physical parameters are listed in Table~\ref{tab:Tab2}. These were used for the following analysis (Section~\ref{sec:Results}). The corresponding synthetic LCs are plotted in Figure~\ref{fig:Lcs}c-d. The nearly flat residuals between synthetic (MCMC analysis) and observed data, suggest, that models describe very well the observed data (Figure~\ref{fig:Lcs}).

\section{Results} \label{sec:Results}
Due to the absence of radial velocities, several indirect methods have been suggested by different authors to extract the absolute physical parameters of the eclipsing binary components. These include semi-major axis-period relations \citep{2016PASA...33...43S}, total mass-period relation \citep{Dimitrov2015}, main-sequence approximation of the massive components, mass-luminosity relations \citep{Sun2020,Ren2021}, period-mass ratio relations \citep{2009CoAst.159..129G}, etc. We adopted the empirical luminosity-mass relation for the primary mass from \cite{2022MNRAS.512.1244C}:
\begin{equation}
\label{eq:1}
\log L_{1}=\log (0.63 \pm 0.04)+(4.8 \pm 0.2)\log M_{1}.    
\end{equation}

\noindent
This relation predicts the observed primary component mass within average fractional mass difference of $\leq 12\%$ for $70\%$ of the systems that were found in literature with masses derived from the analysis of both photometric LCs and radial velocities.
The systems' apparent magnitudes were derived from the LCs obtained from astronomical surveys at quadratures, and the bolometric corrections were adopted from \cite{1998A&A...333..231B}. The extinction was calculated with \textsc{mwdust} \citep{2016ApJ...818..130B,2019ApJ...887...93G} using the distance from {\em Gaia} EDR3 \citep{2021AJ....161..147B} for CRTSJ172718 and the distances from Section~\ref{subsec3:OGLE104},~\ref{subsec4:OGLE-BLG-ECL-00001} for OGLE104 and OGLE012. The luminosity of the primary component was subsequently calculated for each system, and its mass was derived using equation~\ref{eq:1}. Adopting the results from the light curve modeling, the approximate absolute parameters of the eclipsing binary components are estimated and listed in Table~\ref{tab:Tab3}.

For CRTSJ172718, the individual temperatures  were derived from disentangling the value of the binary temperature $T_{\rm sys}$ \citep{2003A&A...404..333Z,2013AJ....146..157C}, using the ratio of relative radii $r_2/r_1$ and the temperature ratio $T_2/T_1$ from the light-curve analysis (see Table~\ref{tab:Tab2}).

\subsection{CRTS J172718.0+431624}
\label{subsec2:CRTSJ172718.0+431624}
CRTS$\_$J172718.0+431624 (V1494 Her), with a period of 0.2250721263 days, $V_{\rm CSS}=14.77$~mag at maximum light and at a distance 542$\pm$ 6 pc, is a shallow ($22\%$) EW with a mass ratio of 0.47$\pm$0.01. Using the reported temperature of the system, namely $4684\pm 183$~K \citep{2019AJ....158..138S}, its components are of K4/K3 spectral type and in good thermal contact. The LCs  presented marked asymmetry in 2016 and interchange between the two light minima between 2016-2022. In the light of new data from ZTF, the final proposed solution from GA optimizer requires a large cool ($\frac{T_{spot}}{T_{1}} = 0.96 $) polar spot on the massive component (star 1) of radius 110$\degr$ at co-latitude 17.5$\degr$ and longitude 110$\degr$. Since spots on the stellar surface can arise and disappear on timescales of days to years due to the magnetic activity particularly in subtype W contact binaries, it is plausible that the spot seen on our 2016 LCs (W subtype) disappeared in the ZTF data and the system appeared as A subtype. Similar switches in relative surface brightnesses of the components were noted on other EW-type systems, TZ Boo \citep{2011AJ....142...99C}, HH UMa \citep{2015ApJ...805...22W} and V410 Aur \citep{2017AJ....154...99L} where the LCs were seen to show (quasi-periodic) exchanges in depths of primary and secondary minima. The model we consider so far uses the mimimum spot coverage necessary to reproduce the observed LC modulations in 2016 data. Doppler imaging technique and a number of spectroscopic activity indicators (the Ca II, H and K lines, H$\alpha$, etc.) should in principle allow a more direct assessment of the activity level. This in combination with LCs of higher quality modelled with genetic algorithms should reveal in future the consistency of the proposed underlying model.

In addition as shown in Figure~\ref{fig:Lcs}e, CSS data (2005-2016) show different maxima in brightness. A similar long term change in mean brightness was also reported by \cite{2017MNRAS.465.4678M} for CSS EW binaries with convective envelopes and by \cite{2018ApJS..238....4P} for CSS eclipsing Algol-type binaries. The latter suggested Applegate mechanism or variable spot regions to explain the cyclic or quasi-cyclic modulation of the observed maximum long-term brightness variability since both mechanisms share the same period of the magnetic cycle of a magnetically active star. This makes this object unexpectedly interesting and accurate times of minimum light observations and period variation analysis are needed to investigate which mechanism(s) may be the underlying cause of these variations.
 
\subsection{OGLE-BLG-ECL-000104} 
\label{subsec3:OGLE104}
 OGLE104 is a contact USPCB with period 0.20075040~d, found by \cite{2015AcA....65...39S} towards the Galactic bulge, at Galactic coordinates $(l,b) = (357.15\degr, -5.12\degr)$. While the probabilistically derived distance from {\em Gaia} EDR3 \citep[$5378 ^{+1507}_{-1240}$~pc;][]{2021AJ....161..147B} supports a location of this object in the bulge, its large uncertainties can affect the derived temperatures of the components, as well as their other calculated absolute parameters.
 
To derive an independent distance estimate along with the extinction, we applied the method of \cite{Ren2021}, who use the period-luminosity relations (PLRs) obtained for different bands by \cite{Chen2018a}. For a given passband, we estimate the maximum absolute magnitudes in $V$, $I$, $K_s$ by adopting the corresponding $\alpha_{\lambda}$ and $\beta_{\lambda}$ coefficients of the PLR and extinction law $A_{\lambda}/A_{V}$ from \cite{Wang2019}, along with the measured maximum  apparent magnitudes $I_{\rm max}=16.628$~mag, $V_{\rm max}=18.256$~mag and $K_{s,{\rm max}}=14.9$~mag from OGLE and VVV data. In this way, we obtain a distance of $1498\pm44$~pc and visual extinction $A_{V}=1.05$~mag. Nevertheless, we have to point out that this relation was derived using $P\geq 0.28$~days systems only \citep{Chen2018a}, and thus requires extrapolation to the USPCB regime. This is the reason why we decided to repeat the previous procedure, but this time using the PLR of \cite{2020Jayasinghe} for the late-type EW binaries separated by period, since this is based on a much larger sample (by $\sim 61$ times) of $\sim 72000$ ASAS-SN contact binaries covering a wider range in distance and $A_{V}$.
Based on a combination of the $VGJHK_s$ bands, the derived distance is 1549~pc and $A_{V}=$1.05~mag.  
To avoid calculating the extinction, we also derive the reddening-free Wesenheit index $W_{JK}$ \citep{1982ApJ...253..575M}, as 
 \begin{equation}
 W_{JK} = K_{s}-0.428 (J-K_{s}), 
 \label{eq:2}
\end{equation}
where we use $R_{JK}=0.428 \pm 0.04$~mag in the VISTA bands \citep{2018A&A...619A...4A}, and adopt $K_{s}$ and color index $(J-K_{s})$ values from \cite{Herpich2021}. We calculate an absolute $W_{JK}=3.671\pm0.062$~mag from the PLR relation of \cite{2020Jayasinghe}, and the distance is estimated from the distance modulus to be 1740~pc. In addition, we also calculate the reddening-free Wesenheit index $W_{VK_{s}}$, as 
 \begin{equation}
 W_{VK_{s}} = K_{s}-0.0846 (V-K_{s}), 
 \label{eq:3}
\end{equation}
where we use $A_{K_{s}}/A_{V}$ from \cite{Wang2019}, mean $K_{s}$ and color index $(V-K_{s})$ from our data, and the corresponding absolute value from the PLR relation of \cite{2020Jayasinghe}. We calculate $W_{VK_s}=3.913\pm0.061$~mag, and a distance of 1538~pc. 

Thus, to run the process of determining the temperature of the primary by conserving the color information, we adopt an average distance of $1580\pm110$~pc, and $A_{V}=1.27$~mag, $A_{I}=0.696$~mag, and $A_{K_{s}}=0.143$~mag from \textsc{mwdust} \citep{2016ApJ...818..130B,2019ApJ...887...93G}. The analysis led to the model presented in Table~\ref{tab:Tab2}. Following the procedure described in Section~\ref{sec:Results} for the adopted distance and extinction values, the photometric solution presented in Table~\ref{tab:Tab2} suggests that OGLE104 is a deep ($80\%$) W-type overcontact binary with K3 components of similar temperature.

In addition, we tried an analysis using the aforementioned distance from {\em Gaia} EDR3 $5378 ^{+1507}_{-1240}$~pc, despite its large uncertainties, and found a model with temperature and relative radii ratios within the errors assigned in Table~\ref{tab:Tab2}, but with a larger fill-out factor ($97\%$). This model solution (not presented here) suggests that OGLE104 may be a deep W-type overcontact binary with K1/K2 components. This discrepancy is to be expected, since a warmer but more distant binary system may also reproduce our photometric data, and highlights the importance of accurate distance determination using future {\em Gaia} parallaxes.

In an attempt to investigate further the effective temperature of the system, we looked at the spectral energy distribution (SED) provided by the VizieR Photometry viewer \citep{Vizier2000}. This revealed the presence of another component within $\leq 1 \arcsec$ that peaks in the infrared ($ 1.02~\mu m$). Although  we do not know if the corresponding source is a foreground source or one that is at the same distance as the system, the presence of an unseen tertiary companion may be revealed by cyclic period variations brought about by the light travel time effect. Such a variation, with a cyclic period of 1400~days, was reported by \cite{2015AcA....65...39S}, based on 12 years of observations. \cite{2020RAA....20..113S} derived a very different model for this system, adopting a solution with partial eclipses, which is not the case as shown in Figure~\ref{fig:Lcs}a. 

\subsection{OGLE-BLG-ECL-000012} \label{subsec4:OGLE-BLG-ECL-00001}
Although OGLE012 was characterized by \cite{2015AcA....65...39S} as a non-contact USPCB towards the Galactic bulge, at Galactic coordinates $(l,b)=(358.31\degr, +2.40\degr)$, with period 0.21894515~days, the recent classification of \cite{2021ApJS..255....1B} and  \cite{2022MNRAS.509.2566M}, as well as our analysis, suggest that it is a contact binary. As the distance from {\em Gaia} EDR3 $2055 ^{+688}_{-606}$~pc \citep{2021AJ....161..147B}, has also large uncertainties, we follow the same steps as with OGLE104 to estimate a distance of $1440\pm100$ pc from the average of four independent methods that use PLR relations and reddening-free Wesenheit indices. We used a larger uncertainty 7$\%$ as for OGLE104. Adopting $A_{V}=1.951$~mag, $A_{I}=1.071$~mag, and $A_{K_{s}}=0.221$~mag from \textsc{mwdust} \citep{2016ApJ...818..130B,2019ApJ...887...93G}, Table~\ref{tab:Tab2} shows that OGLE12 is a shallow  (28$\%$) contact binary of A type with K3/K4 components.

\section{Discussion and Conclusions} \label{sec:Discussion}
Photometric time series in different bands for three USPCBs with total eclipses, were analyzed combining new and archival data from available surveys. Using the genetic algorithm PIKAIA interfaced with PHOEBE/W-D and followed by the MCMC sampling based on W-D, we constructed theoretical models and obtained estimates of the parameters and their uncertainties from the posterior probability distributions of the parameter space for each binary.

\begin{figure}[ht!]
\centering
\epsscale{1.2}
\plotone{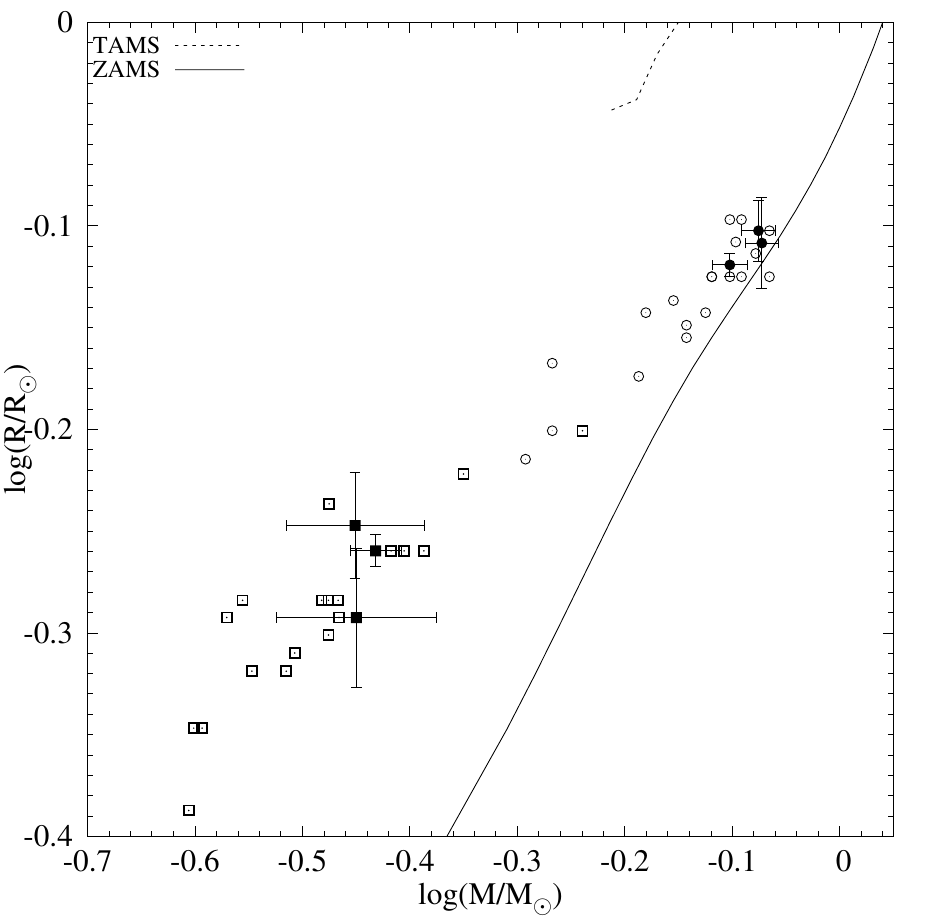}
\caption{ The primary and secondary components (filled circles and squares, respectively) of CRTSJ172718, OGLE104 and OGLE012, compared with primary and secondary components (open circles and squares, respectively) of USPCBs from Table~\ref{tab:biblio},  plotted on the $\log \it{M}-\log \it{R}$ diagram. ZAMS (solid) and TAMS (dotted) lines for solar metallicity, as obtained using the BSE code \citep{Hurley2002}, are overplotted. \label{fig:Evolution}}
\end{figure}

Our review of the literature on well-studied USPCBs with periods below 0.23 days identified  only 28 such systems with spectroscopic mass ratios or/and total eclipses. In most cases, it is because the plethora of new eclipsing binaries with very short periods and  EW type light-curve shapes (contact configuration) discovered by large photometric surveys has not yet been followed up by radial velocity measurements, or even detailed modeling. We collected the parameters of these systems, together with our results, in Table~\ref{tab:biblio}. Among such systems, there are only five spectroscopically confirmed ones. This includes CC Com, the first USPCB identified system with $P=0.2207$~days \citep{1977PASP...89..684R, 2007AJ....133.1977P,2021Zhu}, and the most recent one, 1SWASP J093010.78+533859.5, which is a doubly eclipsing  quintuple system containing a contact binary with $P=0.2277$~days and a detached binary with probably a fifth companion \citep{Lohr2015}. The shortest-period system is the M-dwarf contact binary SDSS J200011.19+003806.5, with $P=0.19856$~days, discovered by \cite{2011ApJ...731...17B} and confirmed spectroscopically by \cite{Davenport2013}. Figure~\ref{fig:Evolution} shows the location of these systems in the $\log M-\log R$ diagram, together with zero-age main sequence (ZAMS) and terminal-age main sequence (TAMS) loci, calculated for solar metallicity using the Binary Star Evolution code \citep[BSE,][]{Hurley2002}.

USPCBs have mass ratios $0.3\leq q\leq0.7$, with a strong concentration around 0.45, and low component masses, from $0.5$ to $0.85 \,M_{\sun}$ for the primary,  $0.25$ to $0.55 \, M_{\sun}$ for the secondary, and in the range from $0.8$ to $1.4 \, M_{\sun}$ for the total mass. All stellar components are K dwarfs (4000-5000~K), except in the cases of 2MASS J02272637+1156494 and SDSS J012119.10-001949.9,  which are of M type. The majority of the systems ($62\%$) is of W subtype (i.e., the more massive component is the cooler). Half of them are included in the catalogue of \cite{2019ApJS..245...34X},  have low metallicities, while the estimated gravitational acceleration of the primary, $\log g_{1}$, is high (see Table~\ref{tab:biblio}), indicating that they are not evolved and may be older stars, in agreement with Figure~7 and Figure~14 of \cite{2020RAA....20..163Q}.
Only four are deep contact systems ($f\geq 50\%$), including OGLE104 studied in this paper, the spectroscopically confirmed GSC 01387–00475\footnote{The low orbital inclination  and the third light contribution of this system make the uncertainties in the fill out factor to be large.}, and the totally eclipsing systems, 1SWASP J074658.62+224448.5 and NSVS 2175434. The remaining ones are shallow contact systems, with $f\leq 20-25\%$.

In order to look at the earlier evolutionary history of USPCBs adopting the detached binary channel, we calculated a set of evolutionary models proposed by \cite{2012Stepien} for low-mass contact binaries (LMCBs), since CC Com and GSC 01387-00475 are common between the LMCBs and USPCBs (although GSC 01387-00475 is not included by the authors as its solution is considered unreliable, due to the very low orbital inclination of the system). Without repeating the details and formulae given in \cite{2012Stepien}, in these models a detached binary composed of two  magnetically active MS stars can evolve to contact through AML via magnetized winds and evolutionary expansion of the initial more massive component or both components. Following a rapid mass exchange until mass ratio reversal, the binary system can reach contact when both components are still on the MS. The evolution path to contact depends on the initial values of the parameters of the progenitor cool detached binary, denoted as $M_{2i}$, $M_{1i}$ and $P_{d}$, where $M_{2i}$ and $M_{1i}$  are the masses of the initial more/less massive component, respectively, and $P_{d}$ is the  initial period in days. In addition, the binary evolution is very sensitive to the mass transfer rate adopted at the first overflow (when the initial massive component of the detached phase fills the Roche lobe for the first time, RLOF) and after mass ratio reversal. The adopted mass transfer rates lie in the range $5.5-7 \times 10^{-9}M_{\sun}~{\rm yr}^{-1}$ \citep{2015A&A...577A.117S}, whereas, for the binary evolution in contact, it falls in the range of $2.55-2.65 \times 10^{-10}M_{\sun}~{\rm yr}^{-1}$\citep{2012Stepien}. Metallicity is also an important  parameter of the initial model, as it determines the MS lifetime and thus the time needed for the massive component to reach RLOF. For a solar composition star, the MS lifetime is about $50\%$ longer, as compared to a star of the same mass but of lower metallicity. 
 
We computed various models with different initial parameters and chose those that can evolve within a Hubble time into a stable contact configuration with a very short period and a comparable value of mass ratio as the observed USPCBs. These are Models~1a-c with initial component masses  $M_{2i}=0.9 M_{\sun}, M_{1i}=0.3M_{\sun}, P_{d}= 2.5d$, low metallicity ($Z=0.001$) and mass transfer rates to RLOF $6.5, 6.85, 7 \times 10^{-9}M_{\sun}~{\rm yr}^{-1}$  respectively. Model 1d has the same initial parameters but solar metallicity and mass transfer rate $5.5 \times 10^{-9}M_{\sun}~{\rm yr}^{-1}$. Model 2a, has initial parameters $M_{2i}=0.8M_{\sun}, M_{1i}=0.3 M_{\sun}, P_{d}= 2.5d$, and  Model 2b has $M_{2i}=0.83 M_{\sun}, M_{1i}=0.33M_{\sun}, P_{d}= 2.5d$. Both have low metallicity ($Z=0.001$), and mass transfer rate to RLOF $5.5\times 10^{-9}M_{\sun}~{\rm yr}^{-1}$.

\begin{figure}[ht!]
\epsscale{1.3}
\plotone{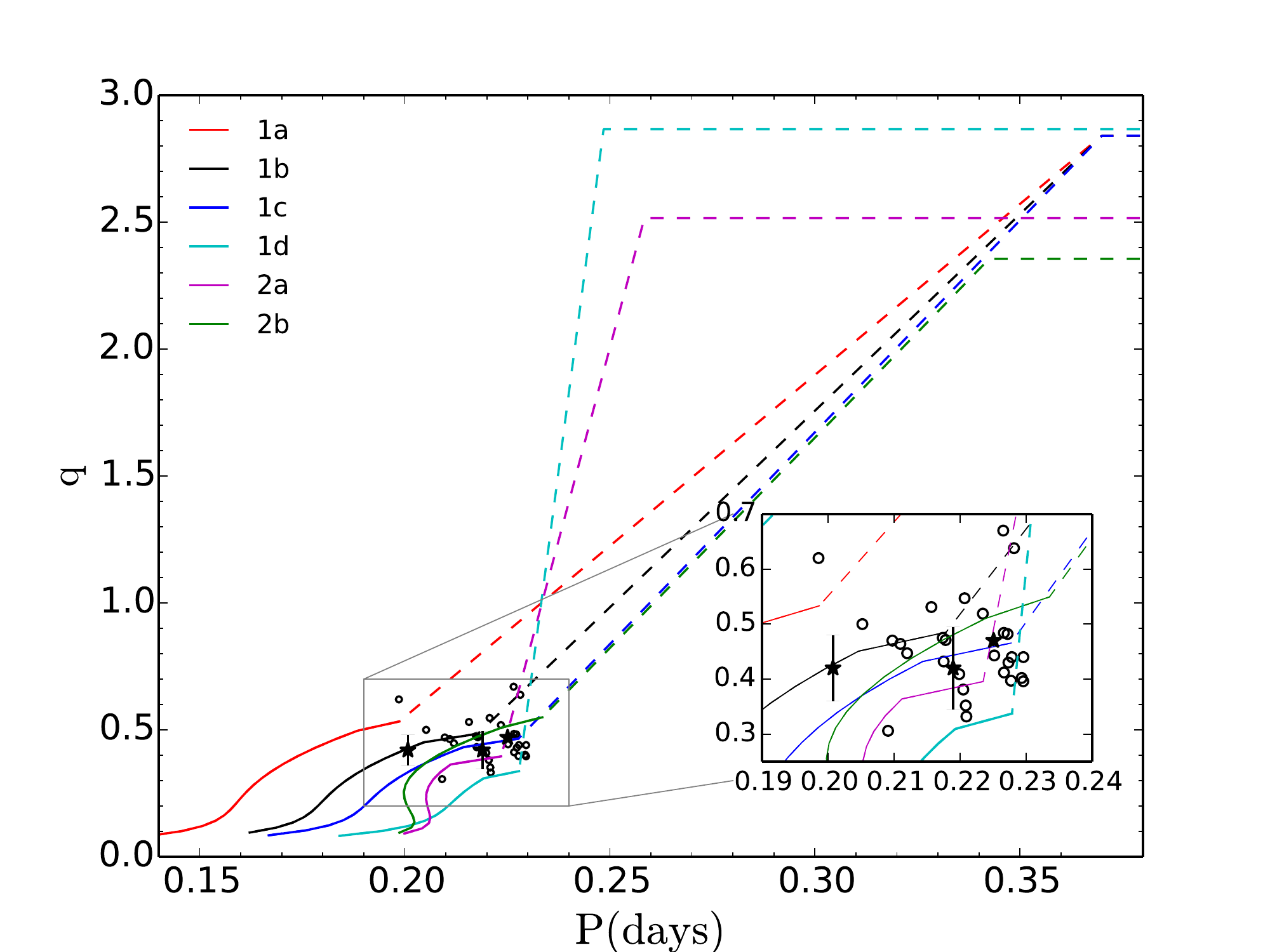}
\plotone{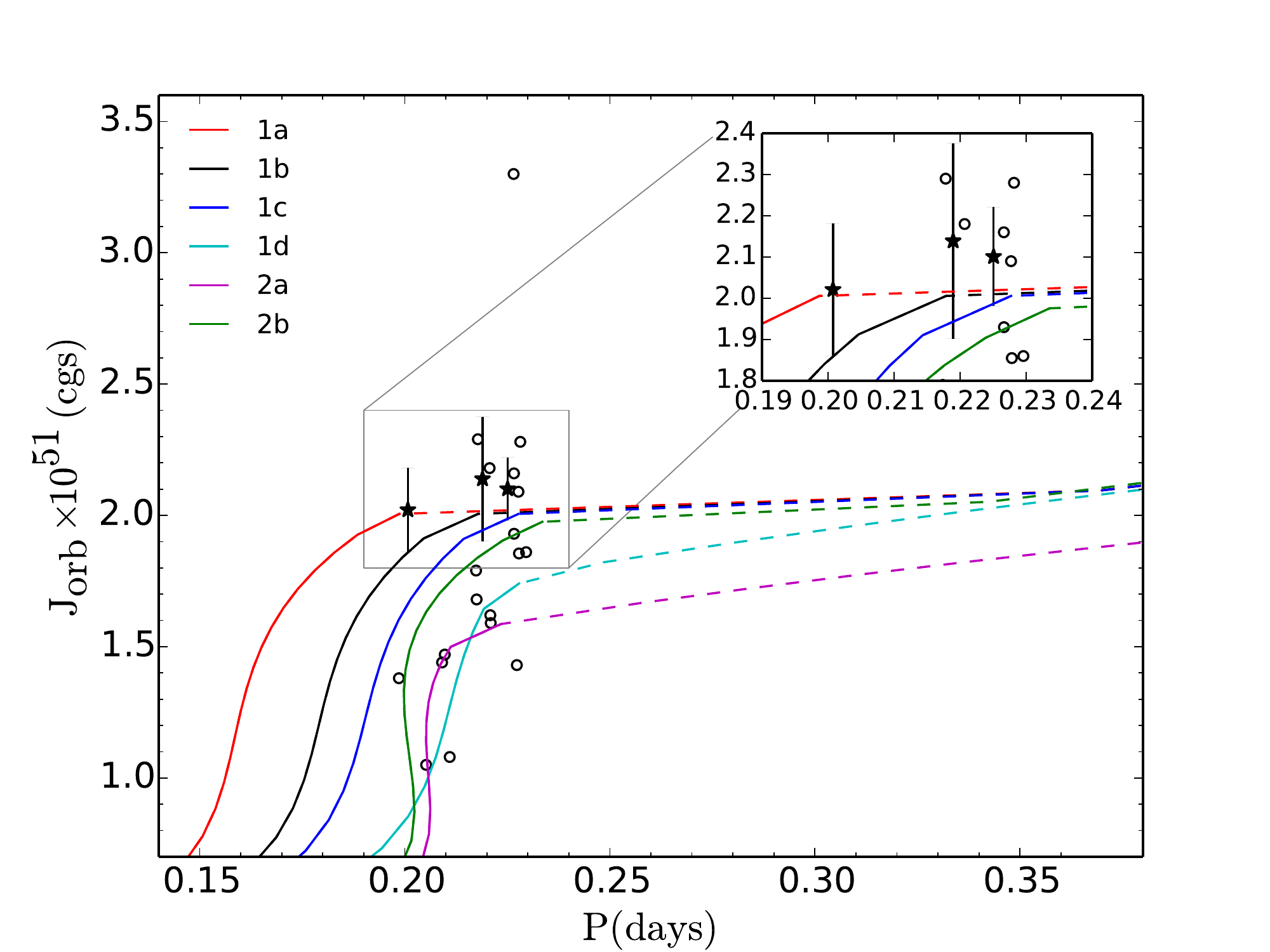}
\caption{ The mass ratios (top) and orbital angular momenta $J_{\rm orb}$ (bottom) of CRTSJ172718, OGLE104 and OGLE012 (filled stars) and USPCBs from Table~\ref{tab:biblio} (open circles) are plotted as a function of the period (in days), and compared with the various evolutionary models (1a-d and 2a-b, coloured lines) described in the text. Contact phases correspond to the solid lines, and pre-contact phases to the dashed lines. \label{fig:Stepien}}
\end{figure}

The comparison of these models with the observed USPCBs is shown in Figure~\ref{fig:Stepien}. In this figure, the top panel gives the observed mass ratio as a function of period of USPCBs of Table~\ref{tab:biblio},  whereas the bottom one shows the orbital angular momentum, $J_{\rm orb}$, as a function of period. $J_{\rm orb}$ (in cgs units) is calculated  by the equation
\begin{equation}
 J_{\rm orb} =1.24 \times 10^{52} M^{5/3}P^{1/3}q (1+q)^{-2}, 
\end{equation}
where $q$ is the mass ratio, $M=M_{1}+M_{2}$ is the total mass of the system (in solar units), and $P$ is the orbital period (in days) from Table~\ref{tab:biblio}. Solid lines correspond to models in contact phase, and dashed lines to earlier phases (ZAMS-RLOF). As our goal is not to determine the unique progenitor of a given USPCB, but to show that a combination of initial parameters can result in their formation, Figure~\ref{fig:Stepien} shows that the predicted values of the mass ratio and the orbital AM in the contact phase of USPCBs are in agreement, within the uncertainties of the assumed mass loss and AML rate. 
The present parameters of the USPCBs depend on the influence of AML and slow mass transfer of the more evolved, low mass component to the present massive. 

In Stepien's model, for an assumed mass loss rate of $10^{-11}M_{\sun}~{\rm yr}^{-1}$, the AML rate brought about by magnetic braking (MB) is given by
\begin{equation}
\frac{dJ_{\rm orb}}{dt}=A\times (R_{1}^2 M_{1}+R_{2}^2 M_{2})/P,  
\end{equation}
where $M_{1,2}, R_{1,2}$ are the masses and radii of both components and $A$ is a numerical coefficient with uncertainty of the order of $50\% $ \citep{2006Stepien}. Thus in the detached binary channel formation of contact binaries, the evolution of the orbital period depends on the assumed MB models and the assumed mass loss by the wind. But even in the same MB model, the evolution depends on the assumed initial mass ratio and period of the system \citep{2014MNRAS.438..859J}. This is caused by the diametrically different way the mass transfer and AML affect the orbital period, %in opposite directions 
and their relative importance determines whether period increases or decreases. Only in the case that a given binary is in a cluster and its age and metallicity are known, can one reproduce well enough its basic parameters (mass, radii, period) within an adopted evolutionary model. \cite{2012Nefs}, taking into account only the orbital evolution, suggested that in order for short-period binaries with primary mass below 0.5~$M_{\sun}$ to exist, a smaller starting period below 1.5~days is needed, so that a contact configuration can be present in less than 12~Gyrs. They showed that binaries with lower $q$ can have even shorter periods at contact.

As an alternative approach for the progenitor parameters of the USPCBs of  Table~\ref{tab:biblio}, we followed the procedure of \cite{2013MNRAS.430.2029Y}, calculating the initial mass ratio using expression (A5) from \cite{2021ApJS..254...10L}. The results of this method  (not presented in Table~\ref{tab:Tab3}) show the initial mass range of the  secondaries as $(1.30-1.64)M_{\sun}$ and that of the primaries as $(0.12-0.50) M_{\sun}$. SDSS J001641-000925, GSC 01387-00476 and SDSS J012119.10-001949.9 have outlier values whereas in the case of 1SWASP J160156.04$+$202821.6 this method is not applicable. These values predict a more massive secondary progenitor and are not in agreement with the above models of Stepien. The main reason is that in the case of USPCBs the assumptions of \cite{2013MNRAS.430.2029Y} concerning the secondary progenitors may not apply or may not be in TAMS phase.

For the four deep contact USPCBs in Table~\ref{tab:biblio}, the presence of a third body may be the alternative/additional process for removing angular momentum from the central eclipsing binary, accelerating the orbital evolution \citep{2007Fabrycky}. Under certain conditions, such a process may play a role in the earlier dynamical evolution of the system through the Kozai mechanism \citep{2001Eggleton,2006Ap&SS.304...75E,2010ASPC..435..169K}. This is indeed the case of the spectroscopically confirmed third component of the system GSC 01387–00475 \citep{2008RucPrib}, but also the possible explanation of the SED of OGLE104 discussed in Section~\ref{subsec3:OGLE104}. Among the remaining USPCBs of Table~\ref{tab:biblio}, SDSS J001641$-$000925 and CC Com also show cyclic period variations, interpreted as an indication of the presence of a third body (denoted by a letter ``$\it{a}$'' in the first column of Table~\ref{tab:biblio}). Three systems, namely CRTS$\_$J014418.3$+$190625, CRTS$\_$J003033.05$+$574347.6, and CRTS$\_$J074350.9$+$451620, show parabolic trends in their period variation studies, of the order of $1-3 \times 10^{-7}d~{\rm yr}^{-1}$, that could also potentially be considered as part of a long-term cyclic variation \citep{2021MNRAS.503.2979P}. Considering the role of orbital period changes in the evolution history of USPCBs, in our future work we are going to investigate/update the period variation of these systems, by collecting many eclipsing times, both new and archival, from photometric surveys, and by applying robust modern and sophisticated optimization methods.

\begin{deluxetable*}{lcCCCCCC}
%\restartappendixnumbering 
\tablecaption{Parameters of CRTSJ172718 and reference stars}\label{tab:Tab1_stars}
\tablewidth{0pt}
\tablehead{
\colhead{Star} &  \colhead{name} & \colhead{\rm RA$_{J2000}$} & \colhead{\rm Dec$_{J2000}$}  & \colhead{J}  & \colhead{H} & \colhead{K$_{s}$}\\
\colhead{} &  \colhead{} & \colhead{(h\,:\,m\,:\,s)}  & \colhead{(\arcdeg : \arcmin : \arcsec)} & \colhead{(mag)}  & \colhead{(mag)} & \colhead{(mag)}
}
\startdata 
Variable (V)	& CRTS$\_$J172718.0$+$431624	&	17:27:18.00		& +43:16:23.72 & 13.18\pm 0.02 & 12.64\pm 0.03 & 12.55\pm 0.03		\\
Comparison (C)	& 2MASS$\_$J17271753$+$4314362	&	17:27:17.54 	& +43:14:36.22 & 12.89\pm 0.02 & 12.55\pm 0.03 &	12.53\pm 0.03	\\
Check (Ch)	    & 2MASS$\_$17271049$+$4316505	&	17:27:10.49	    & +43:16:50.51   & 13.66\pm 0.03 & 13.33\pm 0.03 &	13.28\pm 0.04 \\
\enddata
\tablecomments{Magnitudes were taken from the catalog of 2MASS \citep{2003yCat.2246....0C}.}
\end{deluxetable*}

\begin{deluxetable*}{LCCC}
%\restartappendixnumbering 
\tablecaption{Physical parameters of CRTSJ172718 (ZTF data), OGLE104 and OGLE012 using PIKAIA Genetic Algorithm}\label{tab:Tab2_GA}
\tablewidth{0pt}
%%\tabletypesize{\scriptsize}
\tablehead{
\colhead{ } &  \colhead{CRTS$\_$J172718.0+431624} & \colhead{OGLE-BLG-ECL-000104}& \colhead{OGLE-BLG-ECL-000012}  
}
 
\startdata 
q=\frac{M_{2}}{M_{1}}		&	0.48	\pm	0.03	&	0.44	\pm	0.04	&	0.42	\pm	0.06	\\
T_{r}=\frac{T_{2}}{T_{1}}		&	0.995	\pm	0.004	&	0.997	\pm	0.071	&	0.966	\pm	0.067	\\
R_{r}=\frac{R_{2}}{R_{1}}		&	0.722	\pm	0.021	&	0.728	\pm	0.030	&	0.686	\pm	0.047	\\
r_{1}		&	0.456	\pm	0.009	&	0.510	\pm	0.016	&	0.475	\pm	0.019	\\
r_{2}		&	0.329	\pm	0.009	&	0.371	\pm	0.017	&	0.326	\pm	0.020	\\
\Omega_{12}		&	2.801	\pm	0.061	&	2.553	\pm	0.089	&	2.660	\pm	0.131	\\
i({\degr})	&	88.7	\pm	2.3	&	84.3	\pm	4.0	&	87.1	\pm	3.5	\\
\enddata
\tablecomments{$q,T_{r},R_{r}$ are the mass, temperature and radius ratio of the two components, $r_{1}$ and $r_{2}$ are the mean relative radii, $\Omega_{12}$ is the potential of the components and  $i$ is the orbital inclination.}
\end{deluxetable*}

\begin{deluxetable*}{LLLCC}

%%\tablenum{2}
\tablecaption{Physical parameters of three ultra-short period EWs derived from LC model\label{tab:Tab2}}
\tablewidth{0pt}
%%\tabletypesize{\scriptsize}
\tablehead{
\colhead{ } &  \multicolumn{2}{c}{CRTS$\_$J172718.0+431624} & \multicolumn{1}{c}{OGLE-BLG-ECL-000104} & \multicolumn{1}{c}{OGLE-BLG-ECL-000012}\\ 
\colhead{ } &  \multicolumn{1}{c}{ZTF} & \multicolumn{1}{c}{our data*} & \multicolumn{1}{c}{} & \multicolumn{1}{c}{}  
}
 
\startdata 
{\rm RA_{J2000}} \text{(h\,:\,m\,:\,s)}	&	\multicolumn{2}{c}{17:27:18.00} 	&	\multicolumn{1}{c}{17:59:31.85} 	&	\multicolumn{1}{c}{17:32:12.00}  	\\
{\rm Dec_{J2000}} (\text {$\arcdeg$ : $\arcmin$ : $\arcsec$}) 	&	\multicolumn{2}{c}{43:16:23.72} 	&	\multicolumn{1}{c}{-33:59:16.03} 	&	\multicolumn{1}{c}{-29:05:16.49}  \\
{\rm HJD}_{0} (d)	&	\multicolumn{2}{c}{2457612.48976}	 	&	\multicolumn{1}{c}{2457000.14689}	 	&	\multicolumn{1}{c}{2457000.17111}	 	\\
{\rm Period} (d)	&	\multicolumn{2}{c}{0.22507213} 	&	\multicolumn{1}{c}{0.20075040}	 	&	\multicolumn{1}{c}{0.21894407} 	\\
q=\frac{M_{2}}{M_{1}}		&	0.49^{+0.05}_{-0.04}	&	0.47	^{+	0.01	}_{-	0.01	}	&	 0.42^{+0.07}_{-0.05} 	&		0.42^{+0.08}_{-0.07}\\
T_{r}=\frac{T_{2}}{T_{1}}		&	0.994^{+0.011}_{-0.011}	&	1.000	^{+	0.003	}_{-	0.003	}	&		1.013^{+0.022}_{-0.022} 	&	 0.961^{+0.026}_{-0.027}	\\
R_{r}=\frac{R_{2}}{R_{1}}		&	0.728^{+0.029}_{-0.027}	&	0.72	^{+	0.007	}_{-	0.009	}	&	 0.716^{+0.039}_{-0.034}		&	 0.686^{+0.046}_{-0.034}	\\
r_{1}		&	0.459^{+0.015}_{-0.012}	&	0.463	^{+	0.005	}_{-	0.005	}	&	 0.516^{+0.015}_{-0.018}		&	 0.479^{+0.025}_{-0.020}	\\
r_{2}		&	0.334^{+0.011}_{-0.009}	&	0.334	^{+	0.004	}_{-	0.004	}	&	 0.369^{+0.013}_{-0.012} 	&	 0.329^{+0.016}_{-0.014}	\\
\Omega_{12}		&	2.793^{+0.095}_{-0.105}	&	2.757	^{+	0.029	}_{-	0.032	}	&	 2.511^{+0.129}_{-0.104} 	&	 2.635^{+0.163}_{-0.157}	\\
i({\degr})		&	86.4^{+2.2}_{-2.7}	&	88.4	^{+	1	}_{-	1.3	}	&	 85.4^{+2.7}_{-2.8} 	&	 85.0^{+3.0}_{-2.9}	\\
f(\%)		&	19.9^{+13.5}_{-10.144}	&	22	^{+	5	}_{-	4	}	&	 80^{+11}_{-14} 	&	 28^{+22}_{-16}	\\
\frac{L_{1B}}{L_{Btot}}		&	 \nodata 	&	0.658	^{+	0.005	}_{-	0.005	}	&	 \nodata		&	 \nodata	\\
\frac{L_{1V}}{L_{Vtot}}		&	 \nodata 	&	0.658	^{+	0.005	}_{-	0.005	}	&	 0.648^{+0.025}_{-0.027}		&	 0.739^{+0.031}_{-0.033}	\\
\frac{L_{1R}}{L_{Rtot}}		&	 \nodata 	&	0.656	^{+	0.004	}_{-	0.004	}	&	 \nodata 	&	 \nodata	\\
\frac{L_{1I}}{L_{Itot}}		&	 \nodata 	&	0.657	^{+	0.004	}_{-	0.004	}	&	 0.654^{+0.020}_{-0.022} 	&	 0.716^{+0.026}_{-0.028}	\\
\frac{L_{1K_{s}}}{L_{K_{s}tot}}		&	 \nodata 	&	 \nodata 						&	 0.662^{+0.020}_{-0.023} 	&	 0.707^{+0.032}_{-0.032}	\\
\frac{L_{1g}}{L_{gtot}}		&	0.664^{+0.02}_{-0.021}	&	 \nodata 						&	\nodata	&	\nodata\\
\frac{L_{1r}}{L_{rtot}}		&	0.661^{+0.017}_{-0.018}	&	 \nodata 						&	\nodata	&	\nodata\\
\frac{L_{1i}}{L_{itot}}		&	0.661^{+0.017}_{-0.018}	&	 \nodata 						&	\nodata	&	\nodata\\
\enddata
\tablecomments{$\frac{L_{1j}}{L_{jtot}}$ is the fractional luminosity of the primary component  in filter $j$, and $f=\frac{\Omega-\Omega_{\rm in}}{\Omega_{\rm out}-\Omega_{\rm in}}$ is the fillout factor, where  $\Omega_{\rm in}$ and $\Omega_{\rm out}$ are the modified Kopal potential of the inner and the outer Lagrangian points, respectively. The rest of the parameters as in Table~\ref{tab:Tab2_GA}}
\tablenotetext{*}{\scriptsize Spot parameters on star 1: $\frac{T_{spot}}{T_{1}} = 0.96 $, radius 110$\degr$, co-latitude 17.5$\degr$, longitude 110$\degr$.}
\end{deluxetable*}

\begin{deluxetable*}{LCCC}
%%\tablenum{3}
\tablecaption{Approximate absolute parameters of three ultra-short period EWs \label{tab:Tab3}}
\tablewidth{0pt}
%%\tabletypesize{\scriptsize}
\tablehead{
\colhead{ } & \colhead{CRTS$\_$J172718.0+431624} & \colhead{OGLE-BLG-ECL-000104}& \colhead{OGLE-BLG-ECL-000012}  
}

\startdata  
T_{\rm sys} ({\rm K})	&	4684	\pm	183	&	4911	\pm	65	&	4717	\pm	132	\\
\alpha (R_{\sun})	&	1.64	\pm	0.02	&	1.53	\pm	0.02	&	1.63	\pm	0.03	\\
T_{1} ({\rm K})	&	4684	\pm	183	&	4909	\pm	75	&	4774	\pm	139	\\
T_{2} ({\rm K})	&	4684	\pm	186	&	4915	\pm	131	&	4588	\pm	182	\\
M_{1} (M_{\sun})	&	0.79	\pm	0.03	&	0.84	\pm	0.03	&	0.85	\pm	0.03	\\
M_{2} (M_{\sun})	&	0.37	\pm	0.02	&	0.35	\pm	0.05	&	0.36	\pm	0.06	\\
R_{1} (M_{\sun})	&	0.76	\pm	0.01	&	0.79	\pm	0.03	&	0.78	\pm	0.04	\\
R_{2} (M_{\sun})	&	0.55	\pm	0.01	&	0.57	\pm	0.03	&	0.53	\pm	0.04	\\
L_{1} (L_{\sun})	&	0.21	\pm	0.03	&	0.28	\pm	0.04	&	0.28	\pm	0.04	\\
L_{2} (L_{\sun})	&	0.13	\pm	0.02	&	0.17	\pm	0.03	&	0.11	\pm	0.02	\\
M_{{\rm bol}, 1} ({\rm mag})	&	6.45	\pm	0.17	&	6.12	\pm	0.15	&	6.10	\pm	0.15	\\
M_{{\rm bol}, 2} ({\rm mag})	&	6.96	\pm	0.18	&	6.68	\pm	0.18	&	7.10	\pm	0.23	\\
\enddata

\end{deluxetable*}  
%\epsscale{1.5}

\begin{deluxetable*}{lClCCCCCCCCCc}
%%\tablenum{4}
\tablecaption{Physical parameters and properties of the well-studied USPCBs\label{tab:biblio}}
\tablewidth{0pt}
\tabletypesize{\scriptsize}
\tablehead{
\colhead{Name} &  \colhead{Period} & \colhead{$q$} &\colhead{$M_1$} & \colhead{$T_1$} & \colhead{$T_{2}$} &  \colhead{$R_{1}$} & \colhead{$R_{2}$} & \colhead{$i$} & \colhead{$f$} &  \colhead{$\log{g_{1}}$} & \colhead{$J_{\rm orb}$} & \colhead{Reference}\\
\nocolhead{Common} & \multicolumn1c{(days)} & \nocolhead{Common} & \multicolumn1c{($M_{\sun}$)} & \multicolumn1c{(K)} & \multicolumn1c{(K)} &  \multicolumn1c{($R_{\sun})$} & \multicolumn1c{$R_{\sun}$)} & \multicolumn1c{(\arcdeg)} &  \nocolhead{Common} & \nocolhead{Common} & \multicolumn1c{($\times 10^{51}$cgs)}& \nocolhead{Common}
}

\startdata 
SDSS J001641$-$000925\tablenotemark{\scriptsize a}	&	0.19856	&	0.620\tablenotemark{\scriptsize *}	&	0.54		&	4342	&	3889	&	0.68	&	0.58	&	53.30	&	0.22	&	4.51	&	1.38	&	(1),(15)	\\
OGLE$-$BLG$-$ECL$-$000104	&	0.20075	&	0.420	&	0.84		&	4894	&	4938	&	0.80	&	0.58	&	87.00	&	0.87	&	4.57	&	2.02	&	\textbf{(2)}	\\
SDSS J012119.10$-$001949.9	&	0.20520	&	0.500	&	0.51		&	3840	&	3812	&	0.61	&	0.45	&	83.90	&	0.22	&	4.58	&	1.05	&	(3)	\\
2MASS J21042404$+$0731381	&	0.20909	&	0.306	&	0.81		&	4800	&	4772	&	0.75	&	0.41	&	78.90	&	0.03	&	4.59	&	1.44	&	(4)	\\
NSVS 7179685	&	0.20974	&	0.470	&	0.65		&	3979	&	4100	&	0.67	&	0.48	&	85.50	&	0.19	&	4.60	&	1.47	&	(5)	\\
2MASS J02272637$+$1156494	&	0.21095	&	0.464	&	0.54		&	3759	&	3800	&	0.63	&	0.45	&	82.40	&	0.11	&	4.57	&	1.08	&	(6)	\\
CRTS$\_$J232100.1$+$410736	&	0.21198	&	0.447	&	\nodata		&	4265	&	4431	&	\nodata	&	\nodata	&	82.80	&	0.16	&	\nodata	&	\nodata	&	(7)	\\
CRTS$\_$J053317.3$+$014049	&	0.21565	&	0.531	&	\nodata		&	4246	&	4401	&	\nodata	&	\nodata	&	87.80	&	0.14	&	\nodata	&	\nodata	&	(7)	\\
CRTS$\_$J014418.3$+$190625 & 0.21737 & 0.475 & 0.72 & 4607 &		4819& 0.70 &	0.51 & 82.2 &0.15& 4.61	&	1.79& (18) \\
1SWASP~J080150.03$+$471433.8	&	0.21751	&	0.432	&	0.72		&	4685	&	4696	&	0.71	&	0.49	&	83.80	&	0.14	&	4.59	&	1.68	&	(5)	\\
	%&	0.21752	&	0.451	&	0.82		&	4650	&	4720	&	0.75	&	0.50	&	86.30	&	0.09	&	4.60	&	2.13	&	(4)	\\
GSC 01387$-$00475	&	0.21781	&	0.471\tablenotemark{\scriptsize *}	&	0.835		&	4500	&	4445	&	0.77	&	0.55	&	37.6	&	0.43	&	4.58	&	2.29	&	(8),(16)	\\
OGLE$-$BLG$-$ECL$-$000012	&	0.21894	&	0.420	&	0.85		&	4774	&	4588	&	0.78	&	0.53	&	85.0	&	0.28	&	4.58	&	2.14	&	\textbf{(2)}	\\
1SWASP J151144.56$+$165426.4	&	0.21987	&	0.409	&	\nodata		&	4425	&	4427	&	\nodata	&	\nodata	&	82.90	&	0.19	&	\nodata	&	\nodata	&	(7)	\\
1SWASP J220235.74$+$311909.7	&	0.22048	&	0.381	&	\nodata		&	4967	&	5075	&	\nodata	&	\nodata	&	84.50	&	0.24	&	\nodata	&	\nodata	&	(7)	\\
CC Com\tablenotemark{\scriptsize a}	&	0.22068	&	0.547\tablenotemark{\scriptsize *}	&	0.75		&	3967	&	4148	&	0.72	&	0.55	&	89.30	&	0.22	&	4.60	&	2.18	&	(9),(10)	\\
1SWASP J074658.62$+$224448.5	&	0.22085	&	0.352	&	0.79		&	4543	&	4717	&	0.80	&	0.52	&	81.70	&	0.51	&	4.53	&	1.62	&	(11)	\\
NSVS 2175434	&	0.22095	&	0.332	&	0.81		&	4898	&	4903	&	0.80	&	0.51	&	81.90	&	0.48	&	4.54	&	1.59	&	(11)	\\
CRTS$\_$J020730.1$+$145623	&	0.22344	&	0.519	&	\nodata		&	4239	&	4414	&	\nodata	&	\nodata	&	85.40	&	0.14	&	\nodata	&	\nodata	&	(7)	\\
CRTS$\_$J172718.0$+$431624	&	0.22507	&	0.472	&	0.79		&	4684	&	4684	&	0.76	&	0.55	&	88.42	&	0.22	&	4.58	&	2.10	&	\textbf{(2)}	\\
CRTS$\_$J003244.2$+$244707	&	0.22517	&	0.443	&	\nodata		&	4827	&	4907	&	\nodata	&	\nodata	&	89.90	&	0.23	&	\nodata	&	\nodata	&	(7)	\\
1SWASP J160156.04$+$202821.6	&	0.22653	&	0.670\tablenotemark{\scriptsize *}	&	0.86		&	4500	&	4500	&	0.75	&	0.63	&	79.50	&	0.10	&	4.62	&	3.30	&	(12)	\\
CRTS$\_$J003033.05$+$574347.6& 0.22662 &0.484& 0.79& 5067 &5246 &0.75& 0.55& 82.8&0.24&4.59&2.16&(18) \\
1SWASP J052926.88$+$461147.5	&	0.22664	&	0.412	&	0.80		&	5077	&	5071	&	0.78	&	0.52	&	89.90	&	0.24	&	4.56	&	1.93	&	(13)	\\
CRTS$\_$J151631.0$+$382626	&	0.22719	&	0.482	&	\nodata		&	4741	&	4946	&	\nodata	&	\nodata	&	85.90	&	0.20	&	\nodata	&	\nodata	&	(7)	\\
CRTS$\_$ J074350.9$+$451620 & 0.22732 & 0.430 & 0.66 & 4312 & 4471 & 0.72 & 0.48 & 81.10 & 0.18& 4.57 & 1.43 & (18)\\
1SWASP J093010.78$+$533859.5	&	0.22771	&	0.397\tablenotemark{\scriptsize *}	&	0.86		&	4700	&	4700	&	0.79	&	0.52	&	86.00	&	0.17	&	4.58	&	2.09	&	(14)	\\
1SWASP J212454.61$+$203030.8	&	0.22783	&	0.440	&	0.76		&	4840	&	4810	&	0.75	&	0.52	&	89.01	&	0.13	&	4.57	&	1.85	&	(17)	\\
1SWASP J044132.96$+$440613.7
	&	0.22816	&	0.638	&	0.70		&	4003	&	3858	&	0.73	&	0.60	&	87.67	&	0.25	&	4.56	&	2.28	&	(13)	\\
CRTS$\_$J060855.6$+$622713	&	0.22932	&	0.402	&	\nodata		&	4196	&	4387	&	\nodata	&	\nodata	&	83.30	&	0.15	&	\nodata	&	\nodata	&	(7)	\\
CRTS$\_$J034705.9$+$211309	&	0.22957	&	0.396	&	\nodata		&	4868	&	4774	&	\nodata	&	\nodata	&	80.10	&	0.15	&	\nodata	&	\nodata	&	(7)	\\
1SWASP J050904.45$-$074144.4	&	0.22958	&	0.440	&	0.76		&	4840	&	4933	&	0.75	&	0.52	&	89.51	&	0.14	&	4.57	&	1.86	&	(17)	\\
	&	0.22958	&	0.343	&	\nodata		&	5126	&	5056	&	\nodata	&	\nodata	&	82.70	&	0.15	&	\nodata	&	\nodata	&	(7)	\\
\enddata
\tablenotetext{a}{\scriptsize EWs with cyclic period variation.}
\tablenotetext{*}{\scriptsize Spectroscopic value.}
\tablerefs{\scriptsize (1) \cite{Davenport2013},\textbf{ (2) present study}, (3) \cite{Jiang2015}, (4) \cite{Loukaidou2021arxiv}, (5) \cite{Dimitrov2015}, (6) \cite{Liu2015}, (7) \cite{Li2019}, (8) \cite{2008RucPrib}, (9) \cite{2007AJ....133.1977P},	(10) \cite{2021Zhu}, (11) \cite{Kjurkchieva2018}, (12) \cite{Lohr2014}, (13) \cite{Kjurkchieva2018NewA}, (14) \cite{Lohr2015}, (15) \cite{Qian2015a}, (16) \cite{2021MNRAS.501.2897G}, (17) \cite{2021PASJ...73..132L}, (18) \cite{2020AJ....159..189L}.}
\end{deluxetable*}


\begin{acknowledgments}
This research is co-financed by Greece and the European Union (European Social Fund- ESF) through the Operational Programme «Human Resources Development, Education and Lifelong Learning» in the context of the project “Reinforcement of Postdoctoral Researchers - $2^{nd}$ Cycle” (MIS-5033021), implemented by the State Scholarships Foundation (IKY). Support for MC is provided by the Ministry for the Economy, Development, and Tourism's Millennium Science Initiative through grant ICN12\textunderscore 12009, awarded to the Millennium Institute of Astrophysics (MAS), and by Proyecto Basal ACE210002 and FB210003. C.E.F.L. acknowledges a FAPESP program and the computing facilities at INPE and USP/IAG institutes. EL gratefully acknowledges the support provided by IKY ``Scholarship Programme for PhD candidates in the Greek Universities''.
This research has made use of the VizieR catalogue access tool, CDS, Strasbourg, France.

This paper is dedicated to George Nikolidakis (M.Sc. Hellenic Open University, President of the Hellenic Amateur Astronomical Union) who became one of the casualties during the 2020 pandemic.
\end{acknowledgments}

%% To help institutions obtain information on the effectiveness of their 
%% telescopes the AAS Journals has created a group of keywords for telescope 
%% facilities.
%
%% Following the acknowledgments section, use the following syntax and the
%% \facility{} or \facilities{} macros to list the keywords of facilities used 
%% in the research for the paper.  Each keyword is check against the master 
%% list during copy editing.  Individual instruments can be provided in 
%% parentheses, after the keyword, but they are not verified.
\hspace{5mm}
\facilities{Aristarchos 2.3 m, VISTA 4.1 m} %%OGLE, Catalina Sky Survey

%% Similar to \facility{}, there is the optional \software command to allow 
%% authors a place to specify which programs were used during the creation of 
%% the manuscript. Authors should list each code and include either a
%% citation or url to the code inside ()s when available.

\software{\textsc{PHOEBE-0.31a} \citep{2005ApJ...628..426P}, 
          \textsc{MWDUST} \citep{2016ApJ...818..130B},
          \textsc{PyRAF} \citep{2012ascl.soft07011S},
          \textsc{Astrometry.net} \citep{2010AJ....139.1782L},
          \textsc{EMCEE} \citep{2013PASP..125..306F},
          \textsc{W-D code} \citep{2020ascl.soft04004W},
          \textsc{triangle.py-v0.1.1}\citep{dan_foreman_mackey_2014_11020}
          }

%% Appendix material should be preceded with a single \appendix command.
%% There should be a \section command for each appendix. Mark appendix
%% subsections with the same markup you use in the main body of the paper.

%% Each Appendix (indicated with \section) will be lettered A, B, C, etc.
%% The equation counter will reset when it encounters the \appendix
%% command and will number appendix equations (A1), (A2), etc. The
%% Figure and Table counter will not reset.

%\appendix

%\section{Appendix information}

%% For this sample we use BibTeX plus aasjournals.bst to generate the
%% the bibliography. The sample631.bib file was populated from ADS. To
%% get the citations to show in the compiled file do the following:
%%
%% pdflatex sample631.tex
%% bibtext sample631
%% pdflatex sample631.tex
%% pdflatex sample631.tex

\bibliography{bibliography}{}

%% This command is needed to show the entire author+affiliation list when
%% the collaboration and author truncation commands are used.  It has to
%% go at the end of the manuscript.
%\allauthors

%% Include this line if you are using the \added, \replaced, \deleted
%% commands to see a summary list of all changes at the end of the article.
%\listofchanges

\end{document}